\documentclass[iop,apj,twocolappendix,numberedappendix, appendixfloats,usenatbib]{emulateapj}
\pdfoutput=1 %for arxiv submission to use pdf

\usepackage[breaklinks,colorlinks,citecolor=cyan, linkcolor=magenta]{hyperref} 
\usepackage{apjfonts} 
\usepackage{amsmath}
\usepackage{chngcntr}
\usepackage{enumitem} % for enumeration
\usepackage{booktabs}
\shortauthors{Behmard et al.}

\begin{document}

\received{}
\accepted{}


%\title{Experimentally Determined Binding Energies of Astrophysically Relevant Hydrocarbons
\title{Desorption Kinetics and Binding Energies of Small Hydrocarbons}

%% Use \author, \affil, and the \and command to format author and affiliation 
%% information.  If done correctly the peer review system will be able to
%% automatically put the author and affiliation information from the manuscript
%% and save the corresponding author the trouble of entering it by hand.
%%
%% The \affil should be used to document primary affiliations and the
%% \altaffil should be used for secondary affiliations, titles, or email.

%% Authors with the same affiliation can be grouped in a single

\author{Aida Behmard\altaffilmark{1}}
\affil{$^{1}$Division of Geological and Planetary Sciences, California Institute of Technology, Pasadena, CA 91125, USA}

\author{Edith C. Fayolle\altaffilmark{2,3}, Dawn M. Graninger\altaffilmark{4}, Jennifer B. Bergner\altaffilmark{2}, Rafael Mart\'in-Dom\'enech \altaffilmark{2}, Pavlo Maksyutenko\altaffilmark{2}, Mahesh Rajappan\altaffilmark{2}, and Karin I. \"{O}berg\altaffilmark{2}}
\affil{$^{2}$Harvard-Smithsonian Center for Astrophysics, 60 Garden Street, Cambridge, MA 02138, USA}
\affil{$^{3}$Jet Propulsion Laboratory, California Institute of Technology, 4800 Oak Grove Drive, Pasadena, CA 91109-8099, USA}
\affil{$^{4}$Lawrence Livermore National Laboratory, 7000 East Avenue, Livermore, CA 94550, USA}


%% AASTeX 6.0 supports the ability to suppress the names and affiliations
%% of some authors and displaying them under a "collaboration" banner to
%% minimize the amount of author information that to be printed.  This 
%% should be reserved for articles with an extreme number of authors.  
%% The necessary command are \AuthorCallLimit and \collaborationName.
%% An \AuthorCallLimit=2 call prior to the author list will only show
%% the authors in the first two \author calls.  The \collaborationName
%% defines the collaboration identifier.  Commented examples below.

%\AuthorCallLimit=1
%% Will only show Schwarz & Muench since Schwarz and Muench
%% are in the same \author call. 
%\collaborationName{Friends of AASTeX}
%% will print "The AAS collaboration" after the shortened author list.
%% Note that all the \altaffil information will still be shown so it
%% has to be manually commented out if you do not want it shown.
%%
%% Note that all of these author will be shown in the published article.
%% This feature is meant to be used prior to acceptance to make the
%% front end of a long author article more manageable.

\begin{abstract}
Small hydrocarbons are an important organic reservoir in protostellar and protoplanetary environments. Constraints on desorption temperatures and binding energies of such hydrocarbons are needed for accurate predictions of where these molecules exist in the ice vs. gas-phase during the different stages of star and planet formation. Through a series of temperature programmed desorption (TPD) experiments, we constrain the binding energies of 2 and 3-carbon hydrocarbons (C$_{2}$H$_{2}$ - acetylene, C$_{2}$H$_{4}$ - ethylene, C$_{2}$H$_{6}$ - ethane, C$_{3}$H$_{4}$ - propyne, C$_{3}$H$_{6}$ - propene, and C$_{3}$H$_{8}$ - propane) to 2200--4200 K in the case of pure amorphous ices, to 2400--4400 K on compact amorphous H$_{2}$O, and to 2800--4700 K on porous amorphous H$_{2}$O. The 3-carbon hydrocarbon binding energies are always larger than the 2-carbon hydrocarbon binding energies. Within the 2- and 3-carbon hydrocarbon families, the alkynes (i.e., least-saturated) hydrocarbons exhibit the largest binding energies, while the alkane and alkene binding energies are comparable. Binding energies are $\sim$5--20\% higher on water ice substrates compared to pure ices, which is a small increase compared to what has been measured for other volatile molecules such as CO and N$_{2}$. Thus in the case of hydrocarbons, H$_{2}$O has a less pronounced effect on sublimation front locations (i.e., snowlines) in protoplanetary disks.

\end{abstract}

\maketitle



\section{Introduction} \label{sec:intro}

Simple hydrocarbons are common in protostellar and circumstellar environments \citep{tucker74, betz81, oberg08, pontoppidan14, guzman15}, and may constitute an important reservoir of volatile carbon during planet formation. In Solar System comets, which are thought to preserve the volatile composition of the outer Solar Nebula, hydrocarbon detections include CH$_{4}$, C$_{2}$H$_{2}$, and C$_{2}$H$_{6}$ \citep{mumma96, brooke96, hudson97, kawakita14}. Hydrocarbons have also been detected in Solar Nebula analogs: C$_{2}$H and \emph{c}-C$_{3}$H$_{2}$ at millimeter / sub-millimeter wavelengths, CH$_{4}$ in the near-IR, C$_{2}$H$_{2}$ in the mid and near-IR, and the hydrocarbon radical CH$^{+}$ in the far-IR \citep{lahuis06, gibb07, thi11, qi13_a, gibb13, pontoppidan14, kastner15}. Larger hydrocarbons, such as C$_{3}$H$_{4}$ and C$_{3}$H$_{6}$, are expected to be present in disks and comets since they are frequently observed during early stages of star formation \citep{markwick02}, but have yet to be detected \citep{snyder73}. 

In interstellar and circumstellar environments, hydrocarbons can form through several pathways. During the early stages of cloud formation, when the majority of carbon in the gas phase is in the form of atomic carbon, unsaturated hydrocarbons form efficiently through ion-molecule gas-phase chemistry (e.g., \citealt{agundez13}). These carbon atoms can also adsorb onto grain surfaces where hydrogen addition to adsorbed atomic carbon leads to the formation of CH$_{4}$ (e.g., \citealt{tielens82}). CH$_{4}$ and other small hydrocarbons can then serve as starting points for larger hydrocarbon formation, both through grain surface reactions \citep{moore98, oberg10_a}, and through gas-phase reactions following desorption \citep{charnley04, sakai13}. Warm carbon chain chemistry (WCCC) is initiated by CH$_{4}$ sublimation from icy grain mantles and leads to the formation of long, unsaturated carbon chains \citep{sakai13, graninger16}. In the solid state, hydrocarbon-rich ices are proposed to be the starting point of a rich prebiotic chemistry \citep{kaiser98, bernstein05, hardegree_ullman14}. Gas-phase reactions are proposed as a major source of hydrocarbons in the envelopes of protostars \citep{sakai13, graninger16}.

Hydrocarbons formed during the pre- and protostellar stages of star formation are likely inherited by the protoplanetary disk. In the disk, hydrocarbon chemistry may proceed to produce a new, distinct set of products, though the relative importance of inheritance and in situ chemistry for organic molecules is still debated (e.g., \citealt{cleeves16}). In either case, predicting how hydrocarbons are incorporated into plantestimals and planetary atmospheres requires understanding their division between gas and ice phases throughout protoplanetary disks. This division is set by the locations of hydrocarbon sublimation fronts, and by the ease with which hydrocarbons are entrapped in less volatile ices. Both are important to quantify. This study addresses the former process.

The sublimation front locations of molecular species are dictated by adsorption and desorption kinetics, which are in turn set by their binding energies (e.g., \citealt{viti04, hollenbach09, he16, he17}). Some constraints on binding energies exist for CH$_{4}$, C$_{2}$H$_{2}$, C$_{2}$H$_{6}$, and C$_{3}$H$_{8}$ from different experimental studies \citep{collings04, he16, smith16}. We expand on these through a survey of desorption behavior for all linear 2-3 carbon hydrocarbons. Both pure ice desorption and desorption off H$_{2}$O substrates are investigated. Among these experiments, desorption off compact H$_{2}$O ice is the most relevant for the majority of astrophysical environments \citep{boogert15}. In Section \ref{sec:methods} we present the experimental setup and methods used to characterize hydrocarbon desorption. The experimental results and the binding energies are presented in Section \ref{sec:results} and discussed in Section \ref{sec:discussion}, and some astrophysical implications are presented in Section \ref{sec:astrophys}. 

%Hydrocarbons are also an important organic constituent in comets, the remnants of our protoplanetary disk. This has implications for planetary formation as well because hydrocarbons may be delivered via comets to nascent planetesimals. CH$_{4}$, C$_{2}$H$_{2}$, and C$_{2}$H$_{6}$ have been detected in cometary comae while the third stable 2-carbon hydrocarbon C$_{2}$H$_{4}$ has not been observed in comets, but is involved in one of the dominant formation reactions of C$_{2}$H$_{6}$ (e.g., C$_{2}$H$_{2}$ $\rightarrow$ C$_{2}$H$_{3}$ $\rightarrow$ C$_{2}$H$_{4}$ $\rightarrow$ C$_{2}$H$_{5}$ $\rightarrow$ C$_{2}$H$_{6}$) \citep{mumma96, brooke96, kawakita14, hudson97}. This indicates that C$_{2}$H$_{4}$ likely exists in cometary environments though has evaded detection so far. C$_{2}$H$_{4}$ and C$_{2}$H$_{6}$ content in comets is of particular interest because CH$_{4}$/CO and C$_{2}$H$_{6}$/CO ratios may provide a good proxy of the processed-to-primitive cometary nebular gas ratio, thus placing constraints on the thermal and dynamical evolution of the solar nebula \citep{nuth2000}. Determination of such constraints require accurate and reliable data on C$_{2}$H$_{4}$ and C$_{2}$H$_{6}$ thermal desorption behavior.%

\begin{figure}[t]
		\centering
		\includegraphics[width=0.47\textwidth]{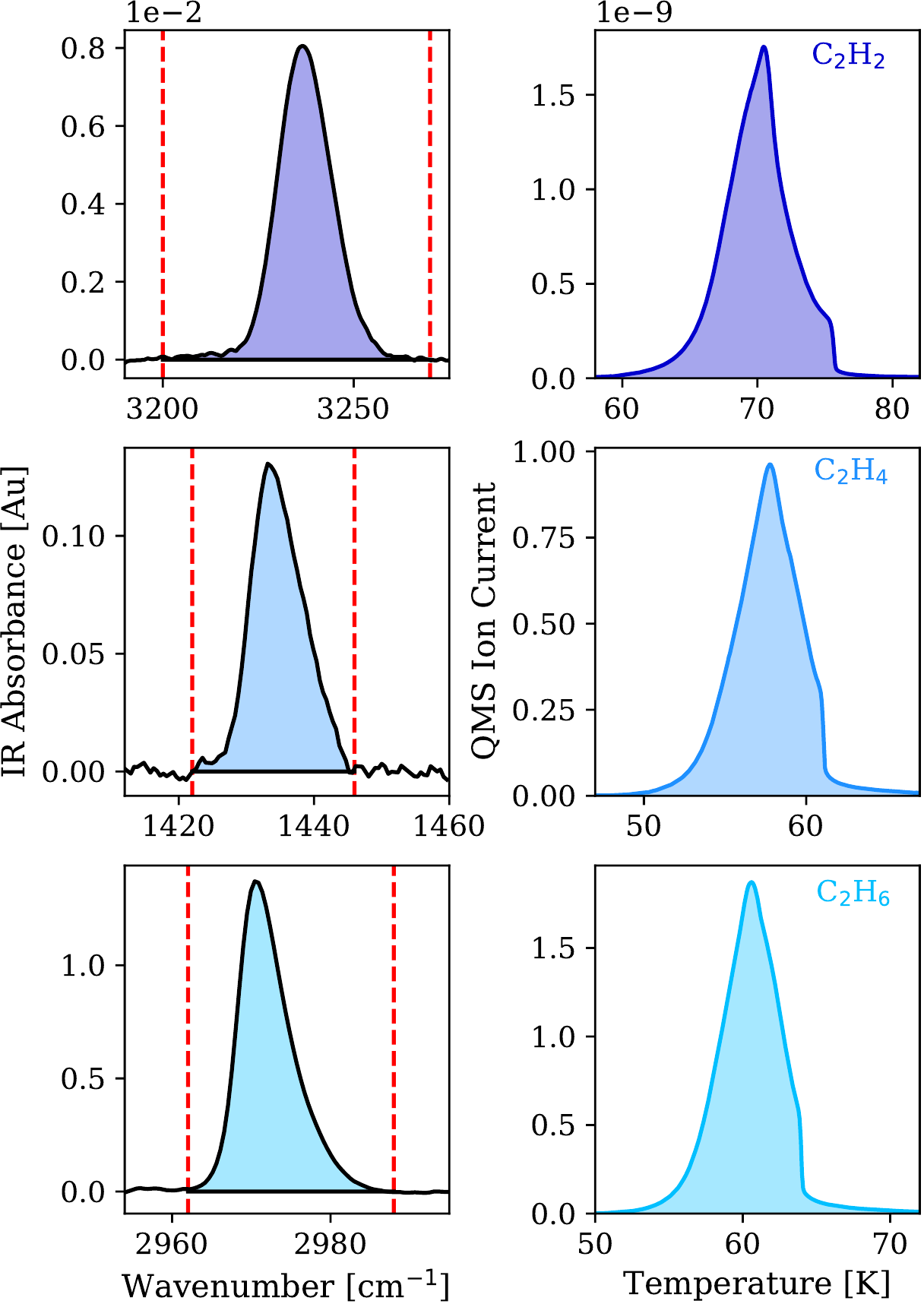}
		\caption{The left column displays the strongest IR features that were used to estimate the ice thicknesses for each pure C$_{2}$H$_{x}$ TPD, with the shaded regions identifying the integrated regions. The right column displays the pure 2-carbon TPDs that were used as references to estimate the thicknesses of C$_{2}$H$_{x}$ TPDs on compact water substrates.}
		\label{fig:figure1}
\end{figure} 

\section{Methods} \label{sec:methods}

\subsection{Experimental Details}
TPD experiments were conducted in an ultra-high vacuum (UHV) chamber described in detail by \citet{lauck15}. The UHV chamber has a base pressure of $<$5 $\times$ $10^{-10}$ Torr at room temperature, dominated by H$_{2}$. Amorphous ices were grown on a 0.75 inch diameter and 2 mm thick CsI window at the center of the chamber that can be cooled to 11 K by a closed-cycle He cryostat. Unless otherwise noted, all ices used in TPD experiments throughout this study are amorphous in structure, which is most relevant for interstellar environments \citep{hagen81,oba09}. Gaseous C$_{2}$H$_{2}$ - acetylene (99.6\% Matheson Trigas) \footnote{The C$_{2}$H$_{2}$ was dissolved in acetone, accounting for its relatively low purity. One should note that we used C$_{2}$H$_{2}$ straight from the bottle without any purification steps because no acetone was detected by quadrupole mass spectrometer analysis during deposition, and no acetone IR features were observed in the FTIR spectrum of deposited C$_{2}$H$_{2}$.} C$_{2}$H$_{4}$ - ethylene (99.99\% Sigma-Aldrich), C$_{2}$H$_{6}$ - ethane (99.99\% Sigma-Aldrich), C$_{3}$H$_{4}$ - propyne ($\geq$ 99\% Sigma-Aldrich), C$_{3}$H$_{6}$ - propene ($\geq$ 99\% Sigma-Aldrich), and C$_{3}$H$_{8}$ - propane (99.97\% Sigma-Aldrich) were deposited through a 4.8 mm diameter pipe with the outlet located 0.8 inches from the CsI window onto the bare 11 K CsI window or onto thick amorphous compact / porous H$_{2}$O ice substrates. The H$_{2}$O was purified beforehand through at least three freeze-pump-thaw cycles using liquid nitrogen. H$_{2}$O substrates were grown by depositing deionized H$_{2}$O at 100 K for compact substrates, followed by cool-down to 11 K, and at 11 K for porous substrates. Deposition temperature influences ice structure (compact v. porous) due to molecular rearrangement resulting from thermal diffusion (e.g., \citealt{bossa12, clements18}). The degree of porosity can be determined from the intensity of the dangling O-H bond spectral feature \citep{devlin95}. However, we were unable to investigate the porosity of our water films via the dangling O-H bond as it is not well detected in our ices (see Fig. \ref{fig:figure9}, left panel; the dangling O-H bond would be visible at $\sim$3600--3700 cm$^{-1}$). This indicates that the porosity of our porous water films is low, but does not indicate a total lack of porosity \citep{raut07, isokoski14}. Laboratory studies also show that the dangling O-H bond cannot be used to investigate porosity quantitatively (e.g., \citealt{bossa14}). To investigate this further, we conducted additional TPD experiments of CO on compact and porous H$_{2}$O substrates, and found that the TPD curve of the porous experiment exhibits a CO desorption peak around the H$_{2}$O desorption temperature, while that of the compact experiment does not (see Fig. \ref{fig:figure9}, right panel). This provides additional evidence that our porous substrates do indeed contain pores. We confirmed that the H$_{2}$O ices were amorphous rather than crystalline in structure from the shapes of the H$_{2}$O IR spectral bands (e.g., \citealt{mastrapa09}).

Following deposition, infrared spectra of ice films were taken using a Fourier transform infrared spectrometer (FTIR, Bruker Vertex 70v) in transmission mode. To produce a single spectrum, 128 scans were averaged over the 4000--600 cm$^{-1}$ spectrometer range at a resolution of 1 cm$^{-1}$. TPD measurements were performed by linearly heating the prepared ices at 2 K min$^{-1}$ and monitoring the desorbing molecules using a quadrupole mass spectrometer (QMS, Pfeiffer QMG 220M1) until complete hydrocarbon desorption. The CsI window temperature was monitored and increased using a temperature controller (LakeShore 355) that operates a heating element situated above the window holder and silicon diode sensors attached onto the window holder. The measured temperature has an estimated accuracy of 2 K and a relative uncertainty of 0.1 K. We obtained TPD plots in desorbing molecules K$^{-1}$ by scaling the main hydrocarbon fragment ion current from the QMS using factors derived from methods explained in Section~\ref{sec:thickness}.

\subsection{Ice Thicknesses} \label{sec:thickness}

This study presents data on pure hydrocarbon ice desorption, and hydrocarbon desorption off compact and porous H$_{2}$O ice. Pure hydrocarbon desorption requires ice thicknesses greater than a few monolayers to ensure that the initial desorption curve is dominated by hydrocarbon-hydrocarbon interactions. In the cases of desorption off H$_{2}$O substrates, ice thicknesses should be in the mono- or submonolayer regime where the kinetics are dominated by hydrocarbon-H$_{2}$O interactions. Ice thickness measurements are required to first ensure that the experiments are carried out in the correct desorption regime, and later to extract binding energies from TPD curves. We also use the estimated ice thicknesses for each experiment to convert the QMS ion current to a desorption rate in units of molecules K$^{-1}$. 

We used three different methods to determine hydrocarbon ice thicknesses: IR spectroscopy (for the pure C$_{2}$H$_{x}$ ices), integrated ion currents from the TPD experiments (for the pure C$_{3}$H$_{x}$ ices and all hydrocarbons desorbing off compact water), and TPD shape characteristics (for all hydrocarbons desorbing off porous water). Errors on ice thicknesses determined from any method other than IR band strengths are taken as 50\%. The ice thicknesses are given in monolayer units with the typical approximation of 1 ML = 10$^{15}$ molecules cm$^{-2}$. However, 1 ML does not always correspond to one molecular layer; porous surfaces are rougher than compact surfaces, allowing them to accommodate more molecules upon a surface area unit. At the same time, porous surfaces contain pores that can trap molecules and inhibit their release via desorption. By comparing thicknesses calculated from IR bands strengths versus using porous experiments as thickness calibrations, we found that these effects roughly cancel each other out, and that the porous H$_{2}$O TPDs provide a reasonable measure of the ion current corresponding to 1 ML.

To determine pure C$_{2}$H$_{x}$ ice thicknesses with IR spectroscopy, we used post-deposition IR spectra and hydrocarbon band strengths. Band strengths relevant for the C$_{2}$H$_{x}$ hydrocarbons in our set are reported in \citet{hudson14a}, \citet{hudson14b}, and \citet{gerakines95} (Fig. \ref{fig:figure1}, left panel). All IR modes and associated band strengths used in this work are reported in Table~\ref{tab:table1}. Thicknesses were calculated from the formula:

\begin{equation} \label{equation1}
%\begin{displaymath}
N_{i} = \frac{ln(10)\int I(\nu)d\nu}{A_{i}}
%\end{displaymath}
\end{equation}

\noindent where $N_{i}$ is the column density (molecule cm$^{-2}$), $\int I(\nu)d\nu$ is the integrated area of the IR band (absorbance units), and $A_{i}$ is the band strength in optical depth units as reported in the literature. Though reported band strength errors are between 0.5--6\%, we adopted an error of 20\% on all band strengths to account for possible differences in temperature and ice structure between our study and those from which the band strengths were extracted. A 20\% error is also consistent with the variance in ice thickness measurements we obtain when different IR bands are chosen for the calculation. 

\begin{figure}[t]
		\centering
		\includegraphics[width=0.48\textwidth]{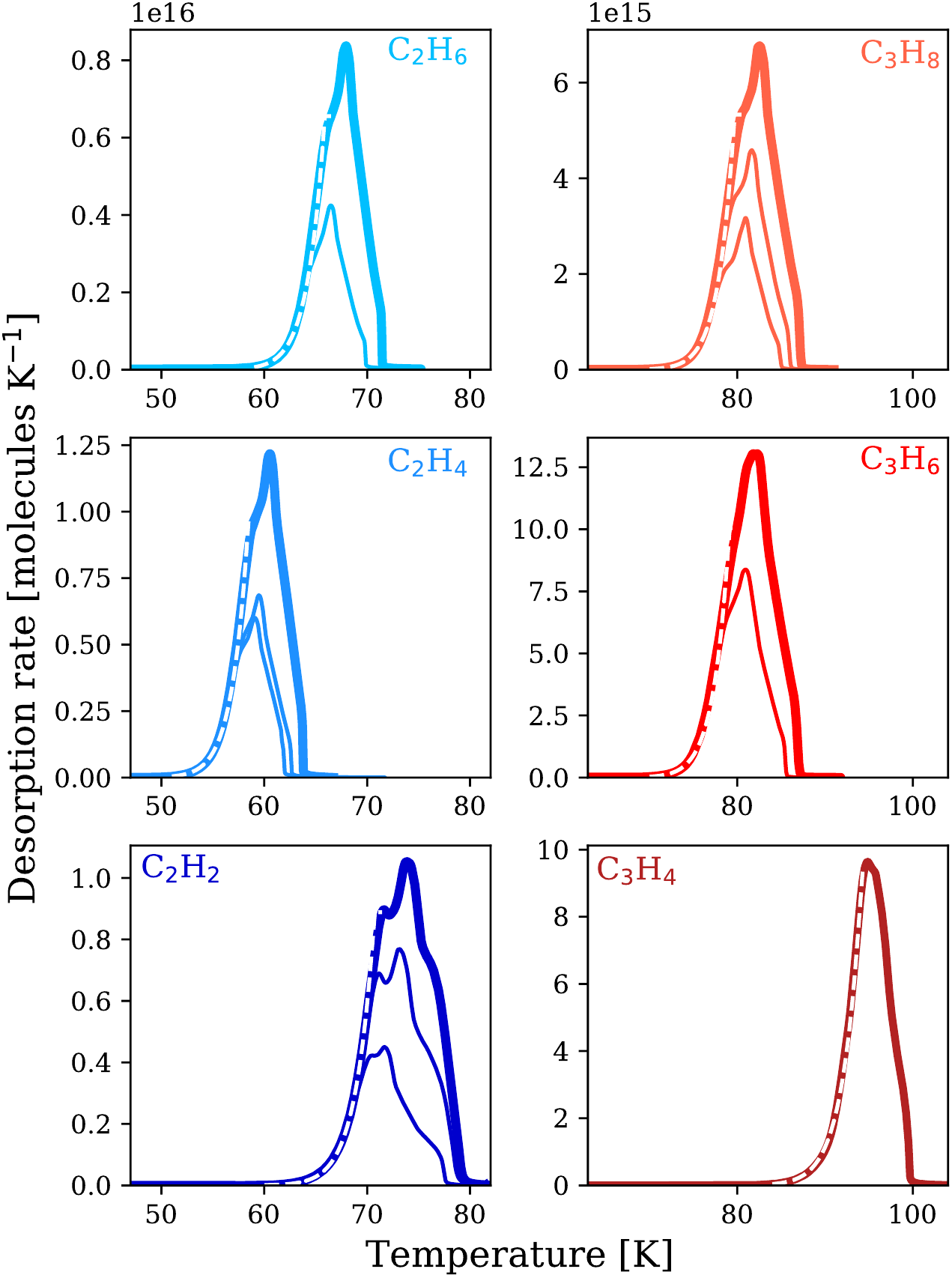}
		\caption{Pure 2-carbon (left) and 3-carbon (right) hydrocarbon TPD curves displayed in solid colored lines, with overlaid white dashed lines representing the fit to obtain the binding energies. When available, TPD runs of thinner ices are overlaid in thinner solid colored lines to demonstrate that the zeroth-order regime was achieved. In the case of C$_{3}$H$_{4}$, the first TPD curve we acquired clearly showed zeroth order kinetics and we therefore did not gather additional, supporting data.}
		\label{fig:figure2}
\end{figure}

Pure C$_{3}$H$_{x}$ ice thicknesses could not be measured with IR spectroscopy due to the lack of C$_{3}$H$_{x}$ band strengths reported in the literature. Instead, they are estimated from their integrated TPD curves using the C$_{3}$H$_{x}$ porous experiments as references, which are assumed to have thicknesses of $\sim$1 ML (justified below). In any case, to ensure that the pure ice experiments (both 2- and 3-carbon) were in the multilayer regime where energies are independent of thickness, we ran a series of TPD experiments of increasing thickness and checked for overlapping leading edges (Fig. \ref{fig:figure2}).

\begin{figure}[t]
		\centering
		\includegraphics[width=0.48\textwidth]{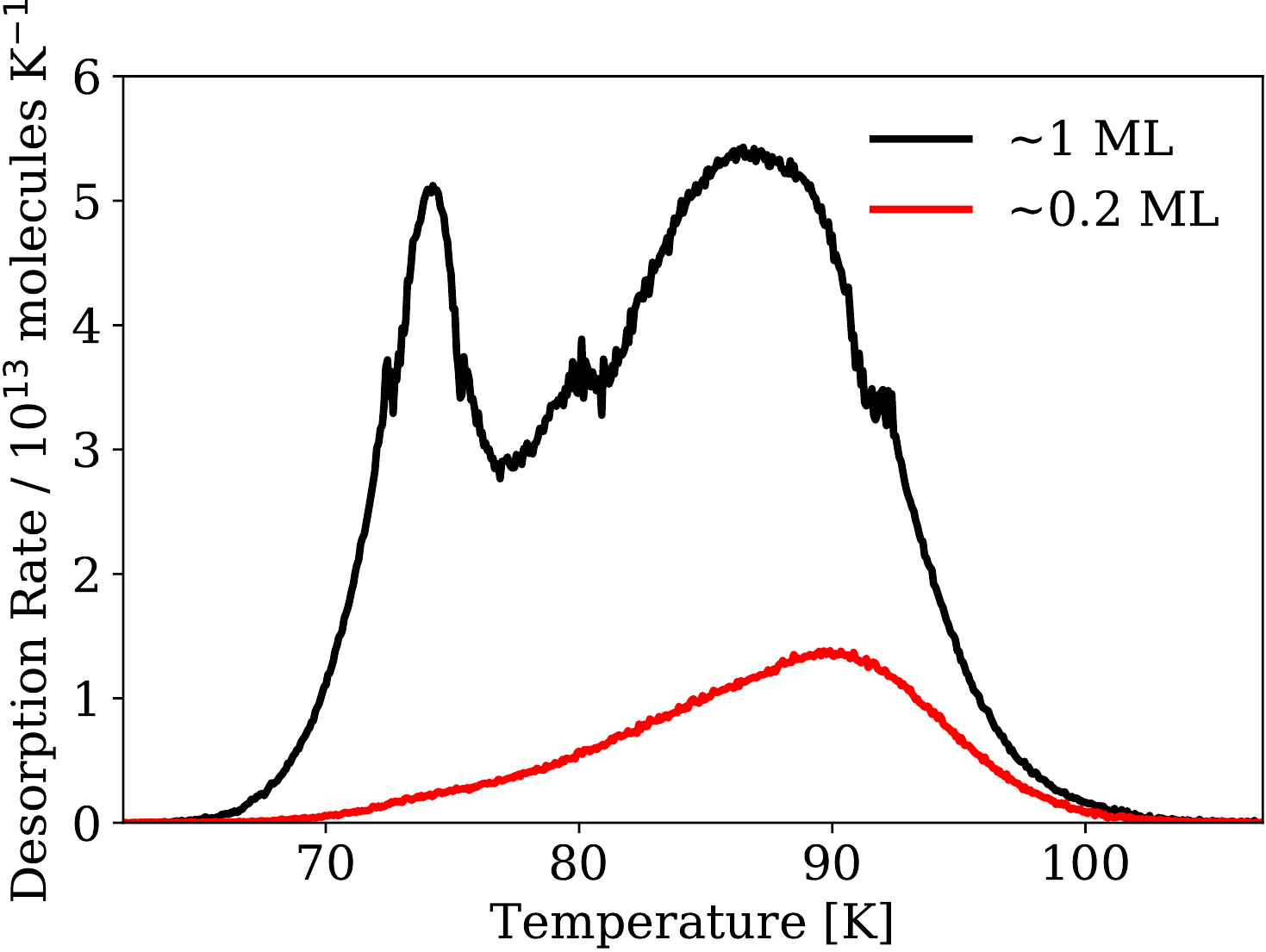}
		\caption{TPD curves of two different C$_{3}$H$_{6}$ coverages on porous H$_{2}$O substrates. The loss of the multilayer peak from the first to the second TPD marks the transition from the multilayer to the submonolayer regime.}
		\label{fig:figure3}
\end{figure}

We estimated the thicknesses of ices (both 2- and 3-carbon hydrocarbons) on porous H$_{2}$O by noting that the TPD curves of all experiments involving porous substrates deviate from the profile expected for pure submonolayer coverages by exhibiting a small multilayer peak, indicating that they are on the verge of reaching the multilayer desorption regime and thus correspond to $\sim$1 ML coverage (Fig. \ref{fig:figure3}).

For experiments of hydrocarbon desorption off compact H$_{2}$O, we aimed to deposit submonolayer coverages of hydrocarbon ice onto a $\sim$50 ML H$_{2}$O ice substrate. The hydrocarbon ice thicknesses could not be verified with IR spectroscopy because IR bands of thin ices are weak and broad in the presence of H$_{2}$O. To obtain ice thicknesses for C$_{2}$H$_{x}$ ices on compact substrates, we compared the integrated areas of their TPD curves (ion current in $A \cdot s$) to those of the pure ice experiments, and multiplied the integrated TPD ratios with the known ice thicknesses of the pure experiments. (Fig. \ref{fig:figure1}, right panel) (e.g., \citealt{doronin15, bertin11}). This method assumes that the QMS signal is proportional to the number of desorbing molecules and that the chamber vacuum pump is evacuating gas at a high speed, both of which have been experimentally verified. Because the C$_{3}$H$_{x}$ band strengths are unknown, we could not use this procedure to estimate the thicknesses of 3-carbon ices on compact substrates. Instead, we used the integrated ion current ratios with the 3-carbon ices on porous substrates as references, as in the case of determining the pure 3-carbon ice thicknesses. 

We chose compact ice experiments with coverages of $\sim$0.2 ML to ensure that we were in the regime where hydrocarbon-H$_{2}$O interactions dominate. Ideally, we wanted the thinnest coverages possible to isolate these interactions, but found that desorption of films thinner than $\sim$0.2 ML did not produce detectable signals in the TPD data. Our choice of $\sim$0.2 ML coverages on compact substrates is further discussed in Section \ref{sec:BEs}.

\begin{deluxetable}{lccr}[h!]
\tablewidth{0.47\textwidth}
\tabletypesize{\footnotesize}
\tablewidth{0pt}
\tablecaption{IR Band Positions and Strengths \label{tab:table1}}
\tablecolumns{4}
\tablehead{
\colhead{Molecule} &
\colhead{IR Mode} &
\colhead{Position} &
\colhead{Band Strength $A_{i}$}  \\
\colhead{} &
\colhead{} &
\colhead{(cm$^{-1}$)} &
\colhead{(cm molecule$^{-1}$)}
}
\startdata
H$_{2}$O           & $\nu_{1}$  & 3280     & 2.0 $\times$ 10$^{-16 \hspace{1mm}}$  \\
C$_{2}$H$_{2}$ & $\nu_{5}$  & 3240     & 2.39 $\times$ 10$^{-17 \hspace{1mm}}$ \\
C$_{2}$H$_{4}$ & $\nu_{7}$  & 1434.3    & 2.24 $\times$ 10$^{-18 \hspace{1mm}}$ \\
C$_{2}$H$_{6}$ & $\nu_{5}$    & 2972.3  & 2.20 $\times$ 10$^{-17 \hspace{1mm}}$  
\enddata
\tablenotetext{}{\textbf{\\Note:} Errors on all band strengths are uniformly set at 20\%. For justification, see Section \ref{sec:thickness}. 
}
\end{deluxetable}

\subsection{Modeling}
To obtain binding energies, we fit the TPD curves with the Polanyi-Wigner equation:

\begin{equation} \label{equation2}
%\begin{displaymath}
-\frac{d\theta}{dT} = \frac{\nu}{\beta} \theta^{n} e^{-E_{b}/T}
%\end{displaymath}
\end{equation}

\noindent where $n$ is the desorption order, $\theta$ is the ice coverage, T is the temperature in K, $\nu$ is a pre-exponential factor in ML$^{(1-n)}$ s$^{-1}$, $\beta$ is the heating rate in K s$^{-1}$, and $E_{b}$ is the binding energy in K. 

For pure ices, we determined the hydrocarbon binding energies by fitting the TPD curves using zeroth-order kinetics ($n$ = 0 in Equation \ref{equation1}). We calculated $E_{b}$ and the pre-exponential factor $\nu$ simultaneously by fitting the logarithm of the desorption rate versus the inverse of the temperature with a straight line. The process is illustrated in Fig. \ref{fig:figure10}.

The hydrocarbon-H$_{2}$O TPDs are fitted with the first-order ($n$ = 1 in Equation \ref{equation1}) form of the Polanyi-Wigner equation as is appropriate for submonolayer desorption where the ice system is characterized by a single binding energy. The non-homogeneous nature of amorphous water ice results in surfaces with a range of binding sites. We therefore fit the submonolayer interaction of the curve with a distribution of binding energies described by a linear combination of first-order desorption kinetics \citep{noble12, collings15, doronin15, fayolle16}. This is accomplished by sampling a range of binding energies in steps of 60--100 K. We used a range of 1800--3700 K for 2-carbon hydrocarbons and 3900--5500 K for 3-carbon hydrocarbons on compact substrates, and a range of 2700--5000 K for 2-carbon hydrocarbons and 3000--5500 K for 3-carbon hydrocarbons on porous substrates. An alternative method for modeling submonolayer desorption is presented in \citet{smith16}, which differs from our approach by modeling the binding energy as a continuous function of coverage. However, we note that our chosen step sizes are well within the binding energy errors and should thus not be a major contributor to the binding energy distribution uncertainties.

We obtain a binding energy distribution from which a mean $E_{b}$ and associated FWHM can be extracted assuming a Gaussian distribution. Pre-exponential factors $\nu$ derived from the multilayer regime calculations are used in the monolayer and submonolayer regime calculations. For more details on the binding energy calculation procedures, see \citet{fayolle16}. Errors on calculations take into account the estimated accuracy of 2 K and relative uncertainty of 0.1 K on the temperature instrument as well as errors on ice thickness (see Section \ref{sec:BEs}).

%Each sampled binding energy is associated with a particular fraction of the initial ice coverage as different coverages result in varying numbers of molecular interactions that collectively reach a maximum at particular binding energy values. 

\begin{figure}[t]
		\centering
		\includegraphics[width=0.48\textwidth]{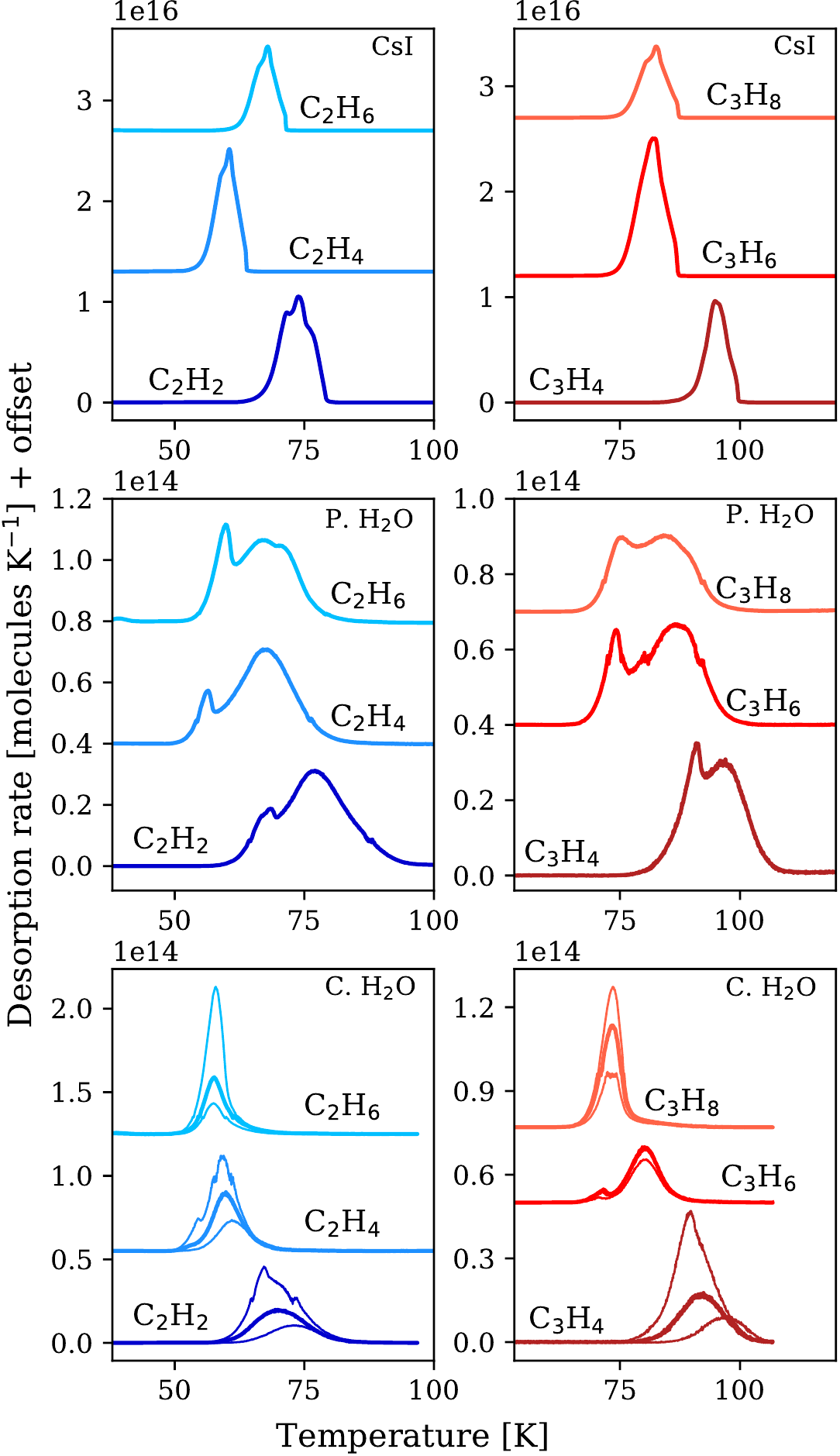}
		\caption{All TPD curves, with TPDs of CsI substrates displayed in the top panels, TPDs of porous H$_{2}$O substrates displayed in the middle panels, and TPDs of compact H$_{2}$O substrates displayed in the bottom panels. The TPDs on compact H$_{2}$O are over-plotted with runs of slightly higher and lower ice coverages for comparison (thinner lines). For all hydrocarbons other than C$_{2}$H$_{6}$ and C$_{3}$H$_{8}$, the higher coverage runs feature multilayer peaks that disappear as coverage decreases. The significantly larger desorption rate scale for the pure TPDs compared to the porous and compact TPDs is due to correspondingly larger coverages / amounts of molecules being desorbed. For zeroth-order desorption, the desorption peak shifts to higher temperatures with increasing coverage, which is why the multilayer peaks for the pure and porous TPDs do not perfectly align.}
		\label{fig:figure4}
\end{figure}

\begin{figure*}[t]
    \centering
    \begin{minipage}{0.516\textwidth}
        \centering
        \includegraphics[width=0.96\textwidth]{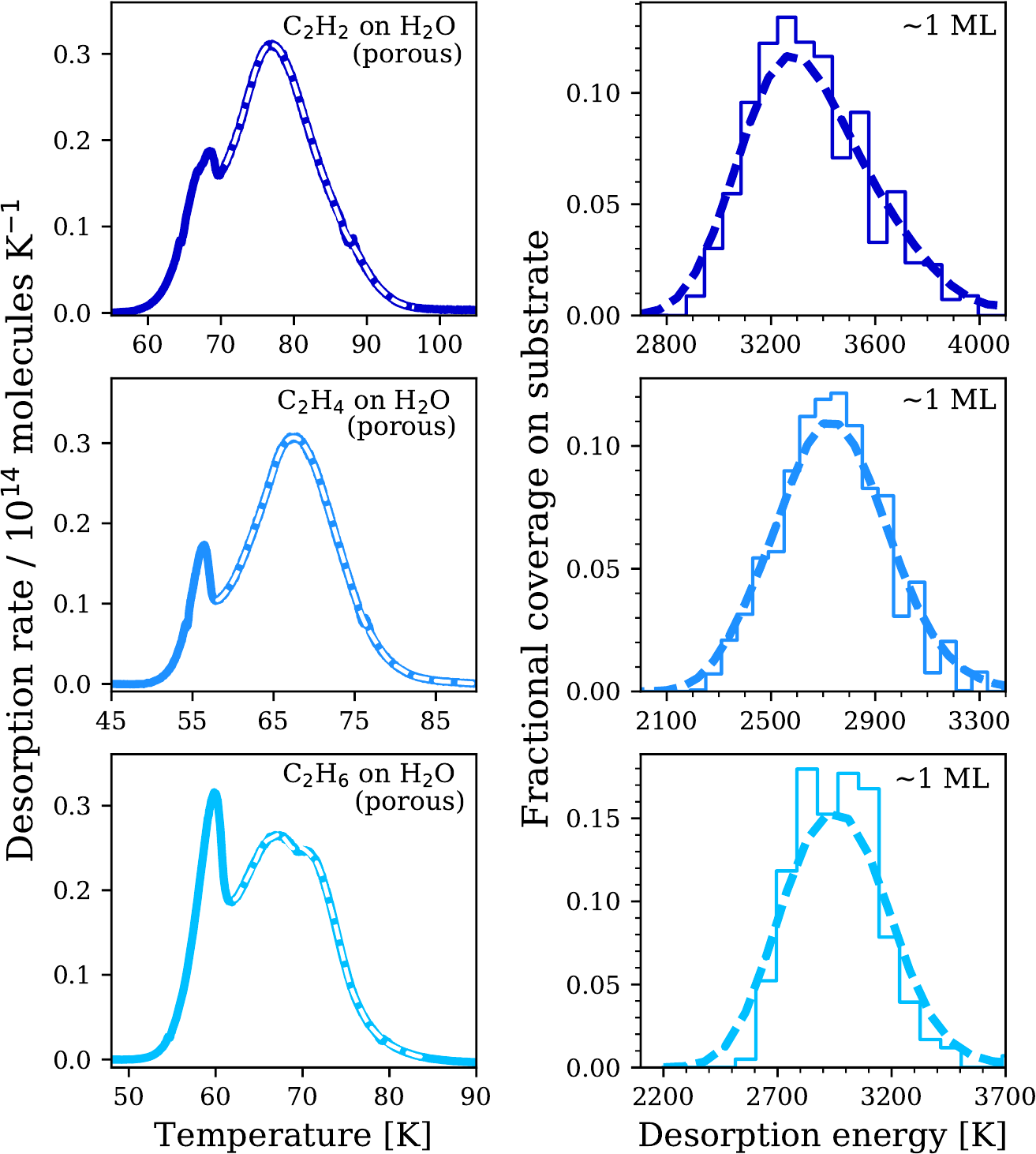} % first figure itself
    \end{minipage}\hfill
    \begin{minipage}{0.48\textwidth}
        \centering
        \includegraphics[width=1.03\textwidth]{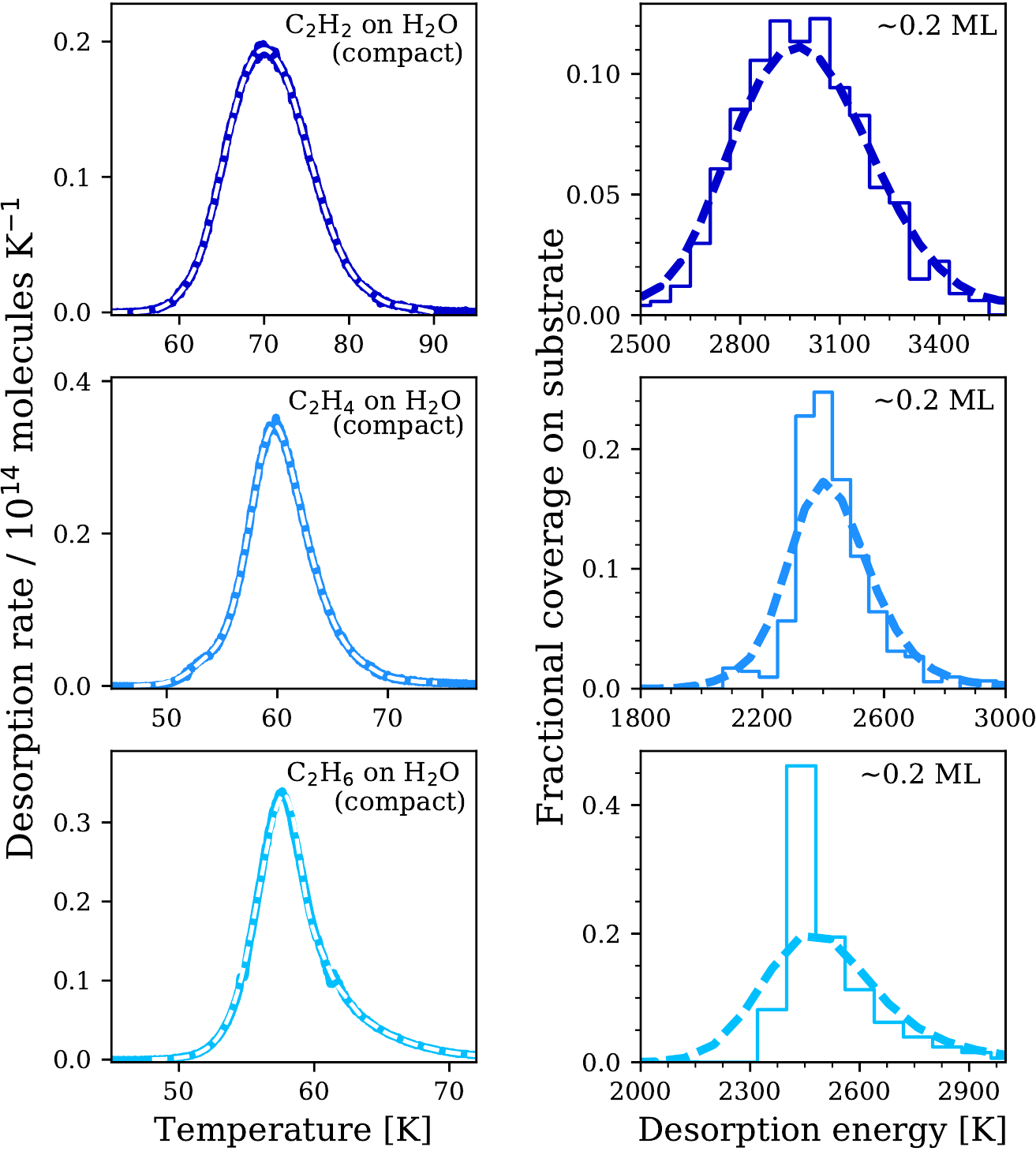} % second figure itself
    \end{minipage}
    \caption{2-carbon hydrocarbon TPD curves on porous (left) and compact (right) H$_{2}$O substrates and corresponding binding energy distributions. TPD curves are displayed in solid colored lines while the white dashed lines represent the fit. The binding energy distributions associated with fractional coverages are shown as the histograms, while the dashed lines represent the smoothed distributions using a Gaussian filter for clarity.}
    \label{fig:figure5}
\end{figure*}

\begin{figure*}[t]
    \centering
    \begin{minipage}{0.48\textwidth}
        \centering
        \includegraphics[width=1.065\textwidth]{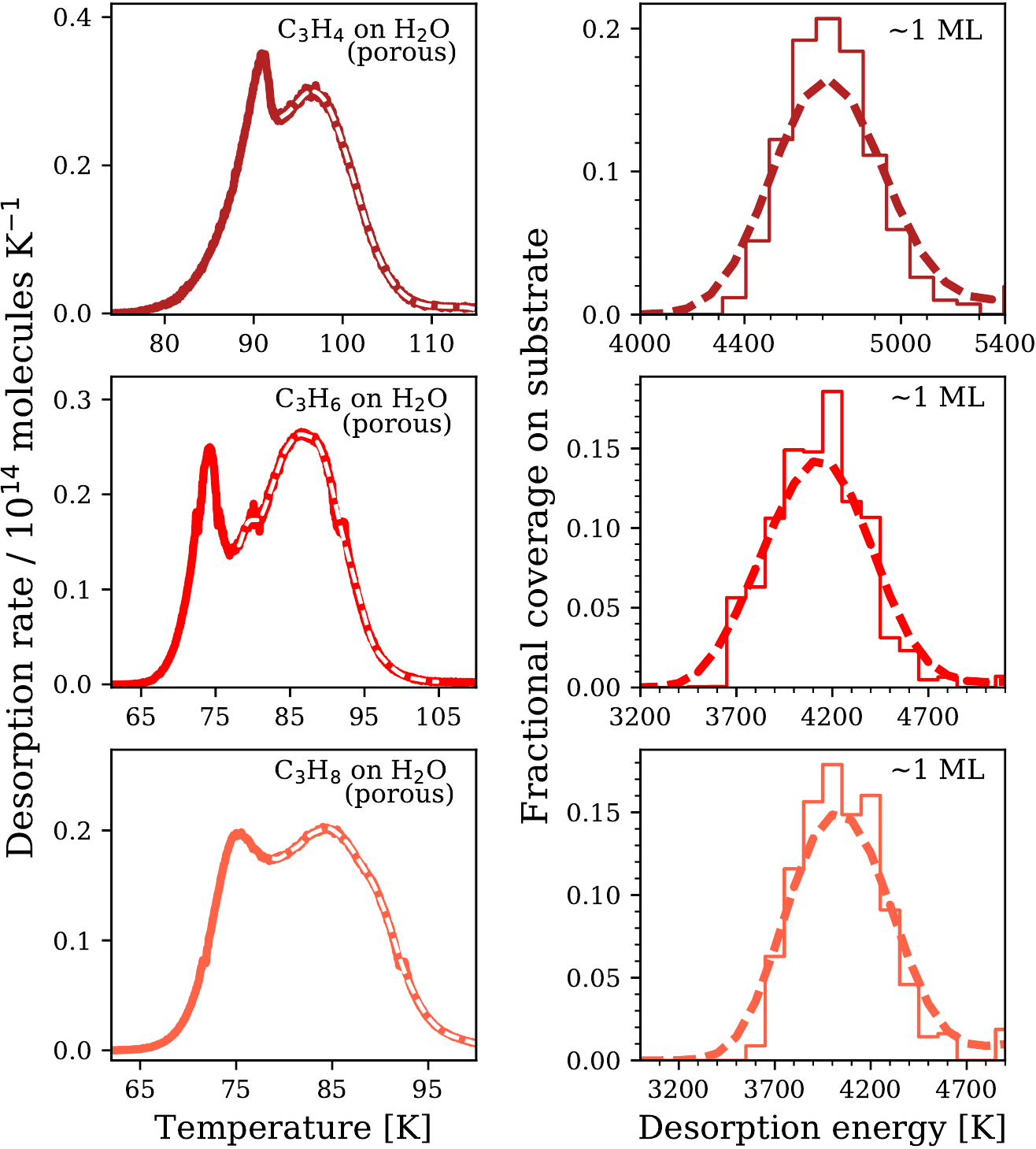} % first figure itself
    \end{minipage}\hfill
    \begin{minipage}{0.5\textwidth}
        \centering
        \includegraphics[width=0.94\textwidth]{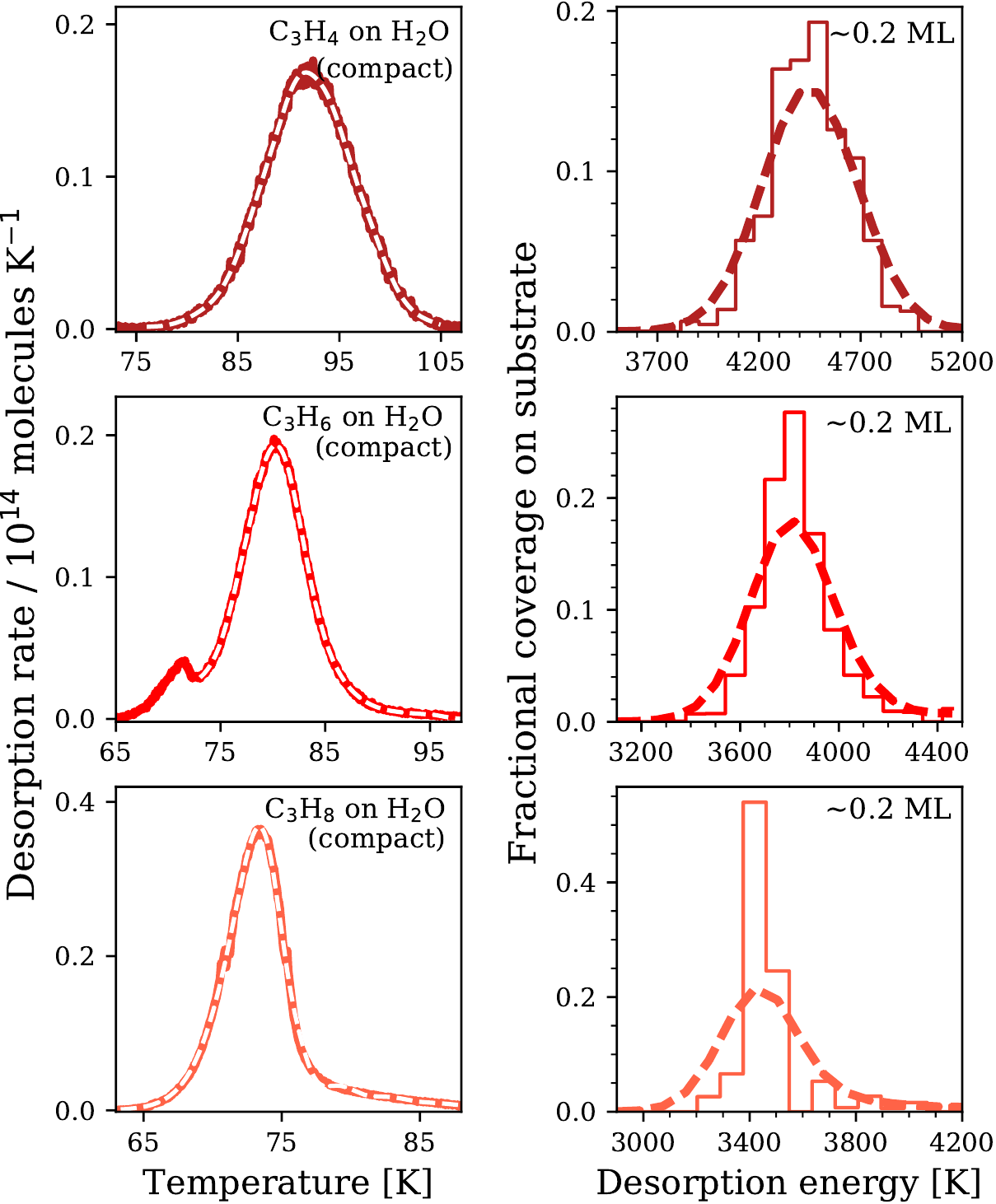} % second figure itself
    \end{minipage}
    \caption{3-carbon hydrocarbon TPD curves on porous (left) and compact (right) H$_{2}$O substrates and corresponding binding energy distributions, similar to Fig.~\ref{fig:figure5}}
    \label{fig:figure6}
\end{figure*}

\section{Results} \label{sec:results}

\subsection{Temperature Programmed Desorption Curves} \label{sec:tpds}
A summary of all experiments is provided in Table~\ref{tab:table2}. TPD curves on CsI, compact H$_{2}$O, and porous H$_{2}$O substrates are displayed in Fig. \ref{fig:figure4}. 

\emph{Pure Hydrocarbon Ices} -- On CsI substrates, 2-carbon hydrocarbons all exhibit lower desorption temperatures than 3-carbon hydrocarbons. Within the 2-carbon hydrocarbon set, C$_{2}$H$_{4}$ exhibits the lowest desorption temperature, followed by C$_{2}$H$_{6}$ and lastly C$_{2}$H$_{2}$. The same desorption temperature trend of alkene, alkane, alkyne is exhibited within the 3-carbon hydrocarbon set, with C$_{3}$H$_{6}$ exhibiting the lowest desorption temperature, followed by C$_{3}$H$_{8}$ and C$_{3}$H$_{4}$ (Fig. \ref{fig:figure2}). 

The TPD curves of 2-carbon hydrocarbons and C$_{3}$H$_{8}$ exhibit fall off from the initial leading edge, resulting in "bump"-like features (Fig. \ref{fig:figure4}). These "bumps" may be due to an amorphous-to-crystalline transition. There are studies that have constrained the temperatures at which amorphous-to-crystalline transitions are expected to occur for 2-carbon hydrocarbons (e.g., \citealt{anderson85, khanna88, zhao88, hudson14a, hudson14b}), and these are generally consistent with the temperature points of the "bumps" in our pure 2-carbon TPD curves (Fig. \ref{fig:figure2}). Whether the bumps observed in the 3-carbon TPDs also coincide with phase changes is unclear due to a lack of experimental data. We did not monitor the ices with the FTIR during warm-up and cannot confirm that there is indeed a crystalline phase transition for any of the experiments.

\emph{Thin Ices on Porous H$_{2}$O Substrates} -- 
All of the hydrocarbons on porous H$_{2}$O present both a multilayer and a submonolayer peak, indicating that the targeted $\sim$1 ML coverage is achieved. As in the pure ice experiments, 2-carbon hydrocarbons have lower desorption temperatures than 3-carbon hydrocarbons. However, the desorption temperature trend changes from that of the pure ice set to alkane, alkene, alkyne for both the 2- and 3-carbon hydrocarbon compact sets. Compared to the compact ice experiments, the porous ice TPD curves appear broader, but as the ice coverages are different no direct comparison is possible. All TPD curves display another large peak near the H$_{2}$O desorption temperature ($\sim$140--160 K) due to release of entrapped molecules (not shown here). %The trapping percentages range from $\sim$40\% to $\sim$60\%.

\emph{Thin Ices on Compact H$_{2}$O Substrates} -- Multiple TPD runs of each hydrocarbon on compact H$_{2}$O substrates were taken to explore the effects of ice coverage around the target coverage of $\sim$0.2 ML.

For the alkanes and alkynes, increasing coverage does not produce the double-peaked desorption curve characteristic of a submonolayer that is more strongly bound than subsequent layers. This is in contrast to the alkenes where a lower temperature desorption peak does appear for the thickest coverages, which can be associated with multilayer desorption (Fig. \ref{fig:figure4}). The lack of such a peak for the alkanes and alkynes indicates that the water-hydrocarbon and hydrocarbon-hydrocarbon interactions are of comparable strengths. This is confirmed by the differences in binding energies for pure alkanes and alkynes, which differ from the binding energies of alkanes and alkynes off compact H$_{2}$O by only $<$10\% (see below in Section \ref{sec:BEs}). Because we are interested in the hydrocarbon-H$_{2}$O interaction, we ran multiple compact TPDs of increasingly thin ices for each molecule until no interaction was visible, then took a slightly higher coverage as the chosen run to ensure that we had isolated the hydrocarbon-H$_{2}$O interaction. We verified that the chosen runs had thicknesses of $\sim$0.2 ML with the methods previously described in Section \ref{sec:thickness}.

As with the pure and porous ices, the 2-carbon hydrocarbons have lower desorption temperatures when compared to the 3-carbon hydrocarbons. As for the porous ice experiments, both the porous 2- and 3-carbon hydrocarbon sets follow the desorption trend of alkane, alkene, alkyne. 

\subsection{Binding Energies} \label{sec:BEs}
TPD curves and the resulting binding energy distributions for the 2- and 3-carbon hydrocarbons on H$_{2}$O substrates are shown in Fig.'s \ref{fig:figure2}, \ref{fig:figure5}, and \ref{fig:figure6}. Derived binding energies and pre-exponential factors are listed in Table~\ref{tab:table2}. We obtain a range in binding energies of 2200--2800 K for pure 2-carbon hydrocarbon ices and a range of 3500--4200 K for pure 3-carbon hydrocarbon ices. Pre-exponential factors that resulted from fitting the pure ice TPD curves range from $\sim$4$\times$10$^{15}$--1$\times$10$^{19}$ s$^{-1}$ and were used to fit the corresponding TPD curves of hydrocarbons on porous and compact H$_{2}$O. While these pre-exponential factors are large, they are not unreasonable given that the hydrocarbons in our set are large; higher pre-exponential factors are correlated with larger molecular sizes because larger chain lengths result in higher rotational entropy, which contributes to the pre-exponential factor calculation \citep{tait05}. In addition, \citet{smith16} report some multilayer C$_{2}$H$_{6}$ and C$_{3}$H$_{8}$ pre-exponential factors in the range 10$^{16}$--10$^{18}$ s$^{-1}$, which is consistent with our results, though also report other pre-exponential factors up to $\sim$3x lower for thinner coverages. Ultimately, our derived pre-exponential factors may be too large, but we must adhere to them because they are derived in conjunction with the binding energies, and we emphasize that they are the proper pre-exponential factors that should be used in chemical models. The uncertainties on the pre-exponential factors are from the 2 K absolute error on the temperature instrument (Table \ref{tab:table2}).

The alkyne and alkene binding energies from desorption off compact H$_{2}$O substrates are $\sim$5--10\% higher than the pure hydrocarbon binding energies. The compact alkane binding energies are consistent with the pure binding energies. Lower binding energies for compact ices than pure ices have been observed in the literature for some species, such as atomic oxygen and O$_{2}$ \citep{he15, noble12}. To check that picking coverages of specifically $\sim$0.2 ML on compact ices did not result in biased binding energy distributions, we derived distributions for additional experiments of $\sim$0.1--0.4 ice coverages on compact ices, and found good agreement between centroids for all coverages (binding energy variation is $\sim$200 K at most) (Table \ref{tab:table2}). For the binding energy distributions, see Fig. \ref{fig:figure11}. Thus, we conclude that varying ice thicknesses does not affect binding energies significantly if the difference is below a monolayer.

Binding energy values increase when moving from compact to porous H$_{2}$O substrate experiments; the binding energies for hydrocarbons desorbing off porous H$_{2}$O are $\sim$5--20\% higher (300--500 K) than those of hydrocarbons desorbing off compact H$_{2}$O. This is expected as diffusion of adsorbate species into substrate pores leads to availability of higher energy binding sites (e.g., \citealt{hornekaer05,  zubkov07, karssemeijer14}). This shift should be considered a lower limit because binding energies for ices on H$_{2}$O substrates decrease with coverage \citep{noble12}, and the porous experiments have substantially higher ice coverages than the compact experiments. 

The only binding energies that exist for this set of hydrocarbons in the literature are reported in \citet{smith16} for C$_{2}$H$_{6}$ and C$_{3}$H$_{8}$ on compact H$_{2}$O. \citet{smith16} report binding energies of 2500 K and 3200 K for C$_{2}$H$_{6}$ and C$_{3}$H$_{8}$ on compact H$_{2}$O ice, respectively, which agree well with the binding energies we obtain from our experiments (Table \ref{tab:table2}). However, more data exist on sublimation enthalpies which can be used to calculate binding energies. A compendium of sublimation enthalpies is reported in \citep{acree16} for a large set of compounds that includes all hydrocarbons used in this study except C$_{3}$H$_{4}$. When these are converted to binding energies, they also agree well with those obtained from our experiments.

The uncertainties on binding energy values are from the errors in ice thickness and the 2 K absolute error on the temperature instrument. The 2 K error is only relevant in the case of pure binding energy values as the FWHM values of the binding energy distributions for the monolayer and submonolayer experiments are always greater than the uncertainties from the 2 K error (Table \ref{tab:table2}). We verified that the ice thickness errors have little effect on the binding energies for the pure ices by taking thickness errors of up to 50\% and noting only negligible shifts in resultant binding energies and derived pre-exponential values. 

%This was verified for ices on H$_{2}$O substrates; varying the assumed thicknesses between 0.1-1 ML for ices on compact H$_{2}$O and 1-2 ML for ices on porous H$_{2}$O produced a negligible effect on the compact and porous binding energy values. We conclude that changing the ice thickness does not have a significant effect on binding energy values for either zeroth or first-order cases. 

\begin{figure*}[t]
	        \centering
		\includegraphics[width=0.95\textwidth]{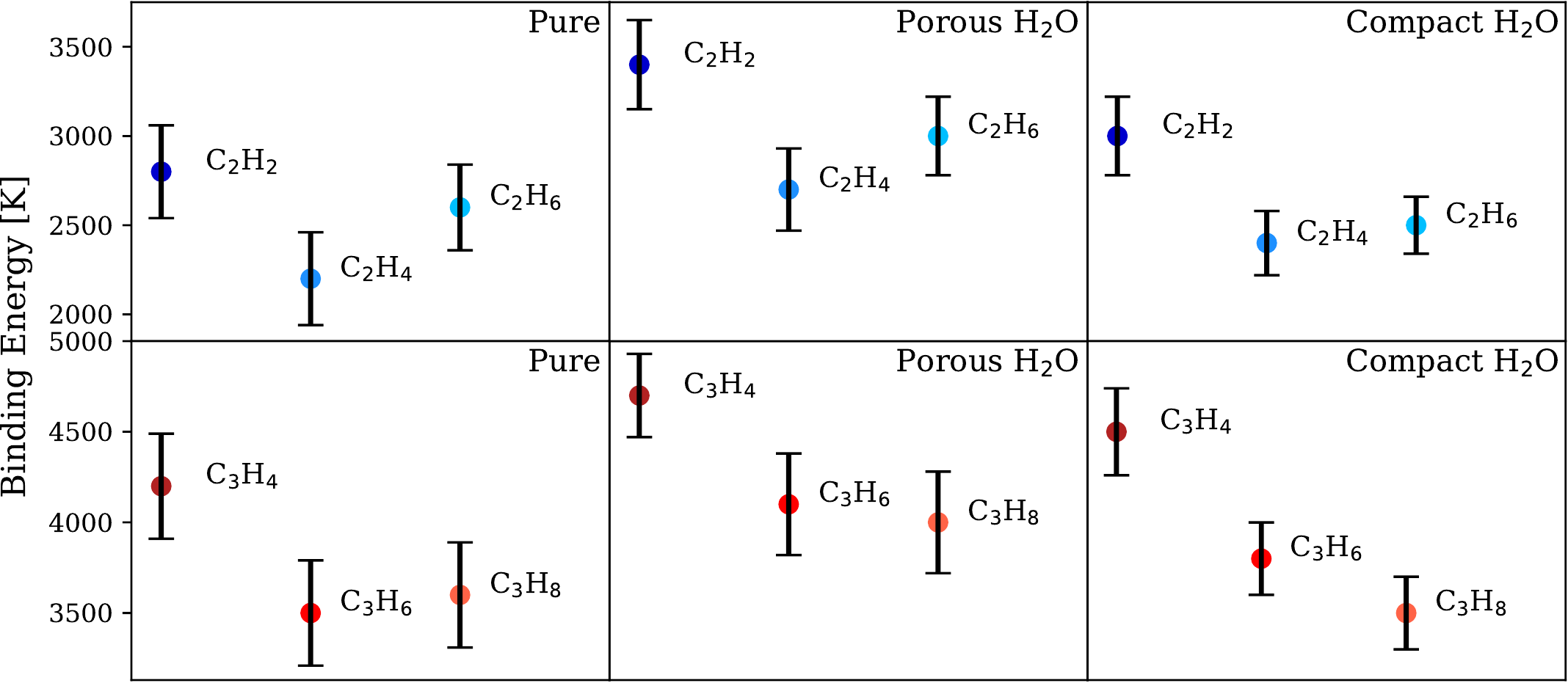}
		\vspace{3mm}
		\caption{Binding energy values of all TPD experiments. Pure ice binding energy are displayed in the left panels, porous H$_{2}$O substrate binding energies are displayed in the middle panels, and compact H$_{2}$O substate binding energies are displayed in the right panels.}
		\label{fig:figure7}
\end{figure*}

\begin{deluxetable*}{cccccc}[h]
\tablewidth{1.0\textwidth}
\tablecaption{Experimental Summary: Binding Energies for Pure Ice Multilayer Regime TPDs and Mean Binding Energies with \\FWHMs for Monolayer and Submonolayer Regime TPDs on H$_{2}$O Substrates (FWHMs indicated in brackets) \label{tab:table2}}
\tablecolumns{6}
\tablehead{
\colhead{Species} &
\colhead{Substrate} &
\colhead{Ice Thickness} &
\colhead{$T_{des}$} &
\colhead{$\nu$} &
\colhead{$E_{b}$}  \\
\colhead{} &
\colhead{} &
\colhead{(ML)} &
\colhead{(K)} &
\colhead{(s$^{-1}$)} & 
\colhead{(K)}
}
\startdata
C$_{2}$H$_{2}$ & CsI Window & 77 $\pm$ 15$^{\hspace{0.5mm}\textrm{(a)}}$ & 74 & 3$^{+17}_{-2.5}$ $\times$ 10$^{16}$ & 2800$^{+200}_{-300}$  \\
\addlinespace[0.12cm]
C$_{2}$H$_{4}$ & CsI Window & 59 $\pm$ 12$^{\hspace{0.5mm}\textrm{(a)}}$  & 61 & 4$^{+36}_{-3.5}$ $\times$ 10$^{15}$ & 2200$^{+200}_{-100}$ \\
\addlinespace[0.12cm]
C$_{2}$H$_{6}$ & CsI Window & 43 $\pm$ 8.6$^{\hspace{0.5mm}\textrm{(a)}}$ & 68 & 6$^{+44}_{-5.2}$ $\times$ 10$^{16}$ & 2600$^{+300}_{-200}$ \\
\addlinespace[0.12cm]
C$_{3}$H$_{4}$ & CsI Window & $\sim50^{\hspace{0.5mm}\textrm{(b)}}$ & 95 & 1$^{+4}_{-0.8}$ $\times$ 10$^{19}$ & 4200$^{+300}_{-200}$ \\
\addlinespace[0.12cm]
C$_{3}$H$_{6}$ & CsI Window & $\sim90^{\hspace{0.5mm}\textrm{(b)}}$ & 82 & 6$^{+34}_{-5.1}$ $\times$ 10$^{18}$ & 3500$^{+300}_{-300}$ \\
\addlinespace[0.12cm]
C$_{3}$H$_{8}$ & CsI Window & $\sim40^{\hspace{0.5mm}\textrm{(b)}}$ & 83 & 4$^{+26}_{-3.4}$ $\times$ 10$^{18}$ & 3600$^{+200}_{-300}$ \\
\addlinespace[0.1cm]
\hline
\addlinespace[0.1cm]
C$_{2}$H$_{2}$ & Compact H$_{2}$O & $\sim$0.2 (0.1, 0.4)$^{\hspace{0.5mm}\textrm{(b)}}$ & 70 & 3$^{+17}_{-2.5}$ $\times$ 10$^{16\hspace{1mm}\textrm{(d)}}$ & 3000$^{+5}_{-5}$ [220] (3100, 3010)\\
\addlinespace[0.12cm]
C$_{2}$H$_{4}$ & Compact H$_{2}$O & $\sim$0.2 (0.1, 0.4)$^{\hspace{0.5mm}\textrm{(b)}}$ & 60 & 4$^{+36}_{-3.5}$ $\times$ 10$^{15\hspace{1mm}\textrm{(d)}}$ & 2400$^{+5}_{-5}$ [160] (2460, 2360)\\
\addlinespace[0.12cm]
C$_{2}$H$_{6}$ & Compact H$_{2}$O & $\sim$0.2 (0.1, 0.4)$^{\hspace{0.5mm}\textrm{(b)}}$  & 58 & 6$^{+44}_{-5.2}$ $\times$ 10$^{16\hspace{1mm}\textrm{(d)}}$ & 2500$^{+10}_{-10}$ [180] (2520, 2480) \\
\addlinespace[0.12cm]
C$_{3}$H$_{4}$ & Compact H$_{2}$O & $\sim$0.2 (0.1, 0.4)$^{\hspace{0.5mm}\textrm{(b)}}$ & 92 & 1$^{+4}_{-0.8}$ $\times$ 10$^{19\hspace{1mm}\textrm{(d)}}$ & 4400$^{+5}_{-5}$ [240] (4600, 4330)\\
\addlinespace[0.12cm]
C$_{3}$H$_{6}$ & Compact H$_{2}$O & $\sim$0.2 (0.1, 0.4)$^{\hspace{0.5mm}\textrm{(b)}}$ & 80 & 6$^{+34}_{-5.1}$ $\times$ 10$^{18\hspace{1mm}\textrm{(d)}}$ & 3800$^{+20}_{-25}$ [200] (3810, 3740)\\
\addlinespace[0.12cm]
C$_{3}$H$_{8}$ & Compact H$_{2}$O & $\sim$0.2 (0.1, 0.4)$^{\hspace{0.5mm}\textrm{(b)}}$ & 73 & 4$^{+26}_{-3.4}$ $\times$ 10$^{18\hspace{1mm}\textrm{(d)}}$ & 3500$^{+5}_{-5}$ [200] (3520, 3490)\\
\addlinespace[0.1cm]
\hline
\addlinespace[0.1cm]
C$_{2}$H$_{2}$ & Porous H$_{2}$O & $\sim$1$^{\hspace{0.5mm}\textrm{(c)}}$ & 77 & 3$^{+17}_{-2.5}$ $\times$ 10$^{16\hspace{1mm}\textrm{(d)}}$ & 3400$^{+50}_{-30}$ [250] \\
\addlinespace[0.12cm]
C$_{2}$H$_{4}$ & Porous H$_{2}$O & $\sim$1$^{\hspace{0.5mm}\textrm{(c)}}$ & 68 & 4$^{+36}_{-3.5}$ $\times$ 10$^{15\hspace{1mm}\textrm{(d)}}$ & 2800$^{+5}_{-5}$ [150] \\
\addlinespace[0.12cm]
C$_{2}$H$_{6}$ & Porous H$_{2}$O & $\sim1^{\hspace{0.5mm}\textrm{(c)}}$ & 67 & 6$^{+44}_{-5.2}$ $\times$ 10$^{16\hspace{1mm}\textrm{(d)}}$ & 3000$^{+40}_{-20}$ [200] \\
\addlinespace[0.12cm]
C$_{3}$H$_{4}$ & Porous H$_{2}$O & $\sim1^{\hspace{0.5mm}\textrm{(c)}}$ & 97 & 1$^{+4}_{-0.8}$ $\times$ 10$^{19\hspace{1mm}\textrm{(d)}}$ & 4700$^{+70}_{-80}$ [230] \\
\addlinespace[0.12cm]
C$_{3}$H$_{6}$ & Porous H$_{2}$O & $\sim1^{\hspace{0.5mm}\textrm{(c)}}$ & 86 & 6$^{+34}_{-5.1}$ $\times$ 10$^{18\hspace{1mm}\textrm{(d)}}$ & 4100$^{+50}_{-60}$ [270] \\
\addlinespace[0.12cm]
C$_{3}$H$_{8}$ & Porous H$_{2}$O & $\sim1^{\hspace{0.5mm}\textrm{(c)}}$ & 84 & 4$^{+26}_{-3.4}$ $\times$ 10$^{18\hspace{1mm}\textrm{(d)}}$ & 4000$^{+60}_{-80}$ [280]  
\enddata
\tablenotetext{(a)}{Derived from IR band strengths, error is taken as 20\%.}
\tablenotetext{(b)}{Derived from integrated ion currents, error is taken as 50\%. Additional binding energies for different submonolayer coverages ($\sim$0.1, 0.4) are provided in the $E_{b}$ column in parentheses.}
\tablenotetext{(c)}{Assumed from TPD shape, error is taken as 50\%.}
\tablenotetext{(d)}{Pre-exponential factors for ices on H$_{2}$O are derived from fitting the corresponding pure ice TPDs.}

\end{deluxetable*}

\section{Discussion} \label{sec:discussion}
We present the binding energies for C$_{2}$H$_{2}$, C$_{2}$H$_{4}$, C$_{2}$H$_{6}$, C$_{3}$H$_{4}$, C$_{3}$H$_{6}$, and C$_{3}$H$_{8}$ from both pure and off porous and compact H$_{2}$O substrates. We observe a clear increase in the binding energies between the 2- and 3-carbon hydrocarbons, and the binding energies for the 2-carbon hydrocarbons are higher than the CH$_{4}$ binding energies in similar ice environments, presenting a clear trend \citep{smith16,he16}.

Within the 2- and 3-carbon families and across all ices, the alkanes and alkenes have similar binding energies while the alkyne binding energies are noticeably higher. This is at odds with the assumption that desorption temperatures / binding energies scale with molecular weight, which is sometimes used in astrochemical simulations when experimental data is lacking \citep{garrod08}. 

For the set of hydrocarbons analyzed, the compact H$_{2}$O-hydrocarbon interactions are only slightly stronger ($\sim$5--10\%) than the hydrocarbon-hydrocarbon interactions. These results differ from those of studies that constrained the effects of compact H$_{2}$O on the binding energies of other molecules, such as CO and N$_{2}$; \citet{fayolle16} determined that the binding energies of CO and N$_{2}$ interactions with compact H$_{2}$O substrates were 30--50\% larger than the binding energies of CO and N$_{2}$ in pure ices when considering similar ice coverages. 

The largest binding energies are achieved on porous H$_{2}$O substrates, which is consistent with previous studies of CO and N$_{2}$ ices on H$_{2}$O \citep{fayolle16, he16}. However, the increase in binding energies ($\sim$10--20\%) from hydrocarbon-hydrocarbon interactions to porous H$_{2}$O-hydrocarbon interactions is again relatively low compared to the $\sim$80\% increase of CO and N$_{2}$ on porous H$_{2}$O \citep{fayolle16}. The relatively low shifts in binding energies for the hydrocarbons on H$_{2}$O can be explained by their hydrophobic nature, which allows attachment to the H$_{2}$O substrate via only weak interactions. 

Binding variations for alkanes, alkenes, and alkynes with H$_{2}$O may arise from differences in hydrocarbon molecular geometry (size, linearity) due to different bonding structures (single, double, or triple bonds) that create different charge densities and orbital hybridizations. Such steric and electronic effects may impact how the hydrocarbons interact with H$_{2}$O. The 2- and 3-carbon family desorption temperature trend of the alkynes exhibiting higher desorption temperatures than the alkenes and alkynes (see Fig. \ref{fig:figure7}) is not obvious, and will require theoretical studies to elucidate. 

\begin{figure*}[t]
	        \centering
		\includegraphics[width=0.98\textwidth]{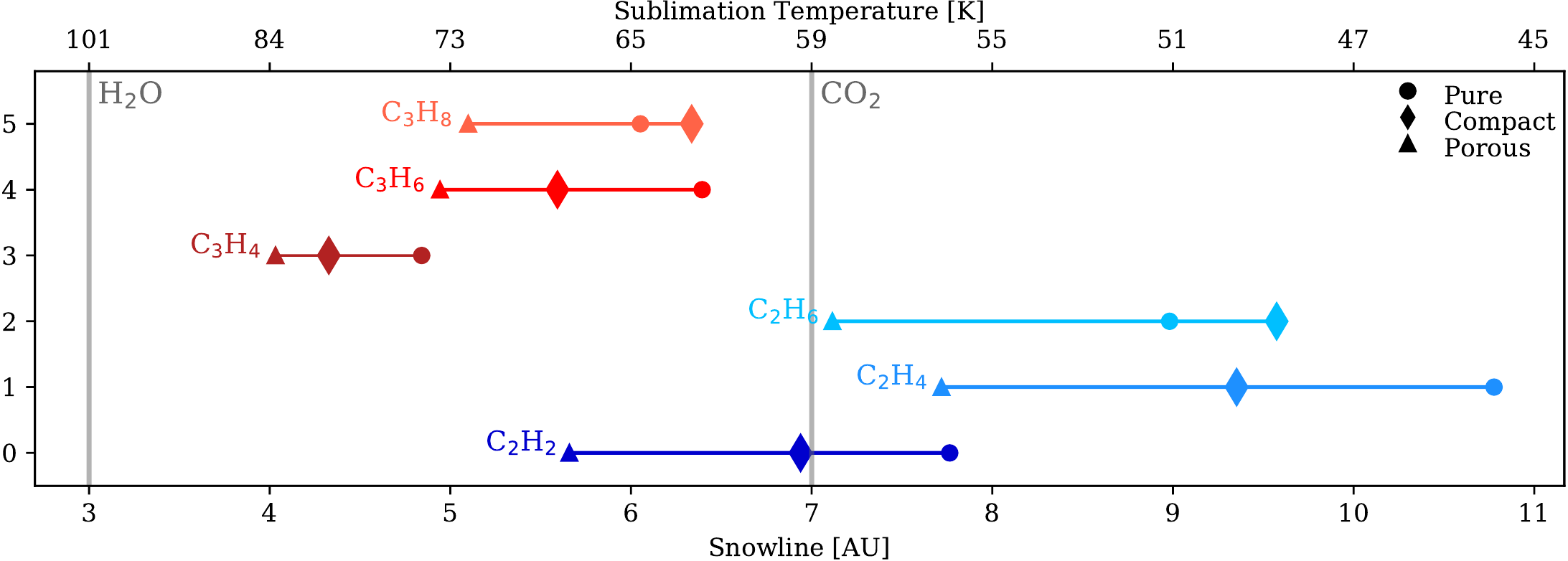}
		\vspace{3mm}
		\caption{Sublimation front locations calculated from binding energies for each hydrocarbon. Binding energies (porous, compact, and pure) are provided above each sublimation front illustration. Sublimation fronts for H$_{2}$O (pure ice) and CO$_{2}$ (in non-porous amorphous H$_{2}$O-dominated ice) are provided for reference. The H$_{2}$O and CO$_{2}$ binding energies used for the sublimation front calculations are taken from \citet{fraser01} and \citet{noble12}, respectively.}
		\label{fig:figure8}
\end{figure*} 

\section{Astrophysical Implications} \label{sec:astrophys}

We use our newly-derived binding energies to estimate the sublimation front locations of all six hydrocarbons in a representative protoplanetary disk, characterized by a disk temperature profile $T = 200 \hspace{1mm} \textrm{K} \times (r/1 \hspace{1mm}\textrm{AU})^{-0.62}$. This is the median disk temperature profile derived from a sample of 24 circumstellar disks in the Taurus-Auriga and Ophiuchus-Scorpius star forming regions \citep{andrews07}. 

Because sublimation front locations are set by sublimation temperatures, we use the prescription from \citet{hollenbach09} to calculate sublimation temperatures from our binding energies, which is derived by equating the flux of adsorbing and desorbing molecules off a grain surface:

\begin{equation} \label{equation3}
%\begin{displaymath}
T_{i} \simeq (E_{i}/k)\Big[\textrm{ln}\Big(\frac{4N_{i}f_{i}\nu_{i}}{n_{i}v_{i}}\Big)\Big]^{-1}
%\end{displaymath}
\end{equation}

\noindent where $T_{i}$ is the sublimation temperature, $E_{i}$ is the binding energy of species $i$, $N_{i} $ is the number of adsorption sites per cm$^{2}$ ($N_{i} \sim 10^{15}$ cites cm$^{2}$), $f_{i}$ is the fraction of the surface adsorption sites that are occupied by species $i$, $n_{i}$ is the gas-phase number density of species $i$, $v_{i}$ is its thermal speed, and $\nu_{i}$ is the vibrational frequency for which we use our derived pre-exponential factors. 

To estimate $f_{i}$, we rely on cometary abundances with respect to H$_{2}$O because there are no hydrocarbon abundances for protoplanetary disks available. The observed abundances of C$_{2}$H$_{2}$ and C$_{2}$H$_{6}$ in cometary ices are 0.2--0.6\% with respect to H$_{2}$O \citep{mumma11}. We further adopt abundances of 0.1\% for C$_{2}$H$_{4}$, and 0.01\% for the 3-carbon hydrocarbons with respect to H$_{2}$O, assuming that cometary abundances decrease by an order of magnitude from CH$_{4}$ (1\% with respect to H$_{2}$O) to C$_{2}$H$_{X}$, and from C$_{2}$H$_{X}$ to C$_{3}$H$_{X}$. As cometary ice is $\sim$80\% H$_{2}$O in composition \citep{delsemme88}, we calculate $f_{i}$ for each hydrocarbon by multiplying the cometary hydrocarbon abundance with respect to water by 0.8 to obtain an estimate of the hydrocarbon surface coverage fraction in icy grain mantles.  

We estimate the gas-phase number density $n_{i}$ by multiplying the number density of atomic hydrogen in protoplanetary disks (which can be regarded as overall density), the H$_{2}$O abundance with respect to hydrogen, and the cometary hydrocarbon abundance with respect to H$_{2}$O. We take the number density of hydrogen in disks to be 10$^{10}$ cm$^{-3}$ in the midplane  (e.g.,  \citealt{oberg11_b}), and the H$_{2}$O abundance to be 10$^{-4}$ per H-atom (e.g., \citealt{boogert15}). 

We then use the disk temperature profile to estimate sublimation front locations from the sublimation temperatures. We find that the 2-carbon hydrocarbons desorb between 6 and 11 AU, or $\sim$70 K and $\sim$50 K, while the 3-carbon hydrocarbons desorb between 4 and 6 AU, or $\sim$80 K and $\sim$70 K. If we limit ourselves to the most likely case of desorption from grains with compact H$_{2}$O mantles, these ranges shrink to 6 and 8 AU for 2-carbon hydrocarbons, and 4 and 6 AU for 3-carbon hydrocarbons (see Fig. \ref{fig:figure8}).

Assuming negligible hydrocarbon entrapment in H$_{2}$O ice, any solid bodies that form within 4 AU will be depleted in small hydrocarbons. This potentially limits their ability to participate in ice chemistry and the organic chemistry of planets that they contribute to. In contrast, planetesimals and planets that form outside 8 AU will be rich in hydrocarbons. This includes comets, which can migrate and deliver material to other bodies throughout the disk. However, entrapment is certainly possible, and future studies on entrapment efficiencies of different hydrocarbons are needed to obtain a complete picture of the distributions of small hydrocarbons during planet formation.

\section{Conclusions}

In this study, we obtained binding energies C$_{2}$H$_{2}$, C$_{2}$H$_{4}$, C$_{2}$H$_{6}$, C$_{3}$H$_{4}$, C$_{3}$H$_{6}$, and C$_{3}$H$_{8}$ in both pure ices and off porous and compact H$_{2}$O. We found:

\begin{enumerate}
\item The binding energies of pure 2- and 3-carbon amorphous ices range from 2200--2800 K and 3500--4200 K, respectively.

\item In the submonolayer regime, the binding energies of 2- and 3-carbon amorphous ices off compact H$_{2}$O substrates range from 2400--3000 K and 3500--4400 K, respectively. Off porous H$_{2}$O substrates, the binding energies of 2- and 3-carbon amorphous ices range from 2800--3400 K and 4000--4700 K, respectively. These porous binding energies are $\sim$10--20\% larger than the pure ice binding energies.

\item Within the 2- and 3-carbon hydrocarbon sets, the alkynes (i.e., least-saturated) hydrocarbons exhibit the largest binding energies.

\end{enumerate}

From these results, we can draw the following conclusions:

\begin{enumerate}
\item There is a relatively small difference in binding energies between pure hydrocarbon ices and hydrocarbon ices desorbing off H$_{2}$O compared to what has been measured for other volatile species (CO, N$_{2}$). This implies that H$_{2}$O has a small influence on the snowline locations of these hydrocarbons in protoplanetary disks.

\item Though the alkynes (C$_{2}$H$_{2}$ and C$_{3}$H$_{4}$) are the smallest molecules within the 2- and 3-carbon hydrocarbon sets, they exhibit higher binding energies, demonstrating that molecular size does not necessarily correlate with larger desorption temperatures / binding energies within molecular families.

\end{enumerate}

\acknowledgments

AB acknowledges funding from the Origins of Life Initiative at Harvard. E.C.F.'s contribution was partly carried out at the Jet Propulsion Laboratory, California Institute of Technology, under a contract with the National Aeronautics and Space Administration. KI\"{O} acknowledges funding from the Simons Collaboration on the Origins of Life Investigator award \#321183.

\newpage
\appendix

\section{Supplementary Figures}
\counterwithin{figure}{section}

\begin{figure*}[t]
	        \centering
    \begin{minipage}{0.48\textwidth}
        \centering
        \includegraphics[width=1.035\textwidth]{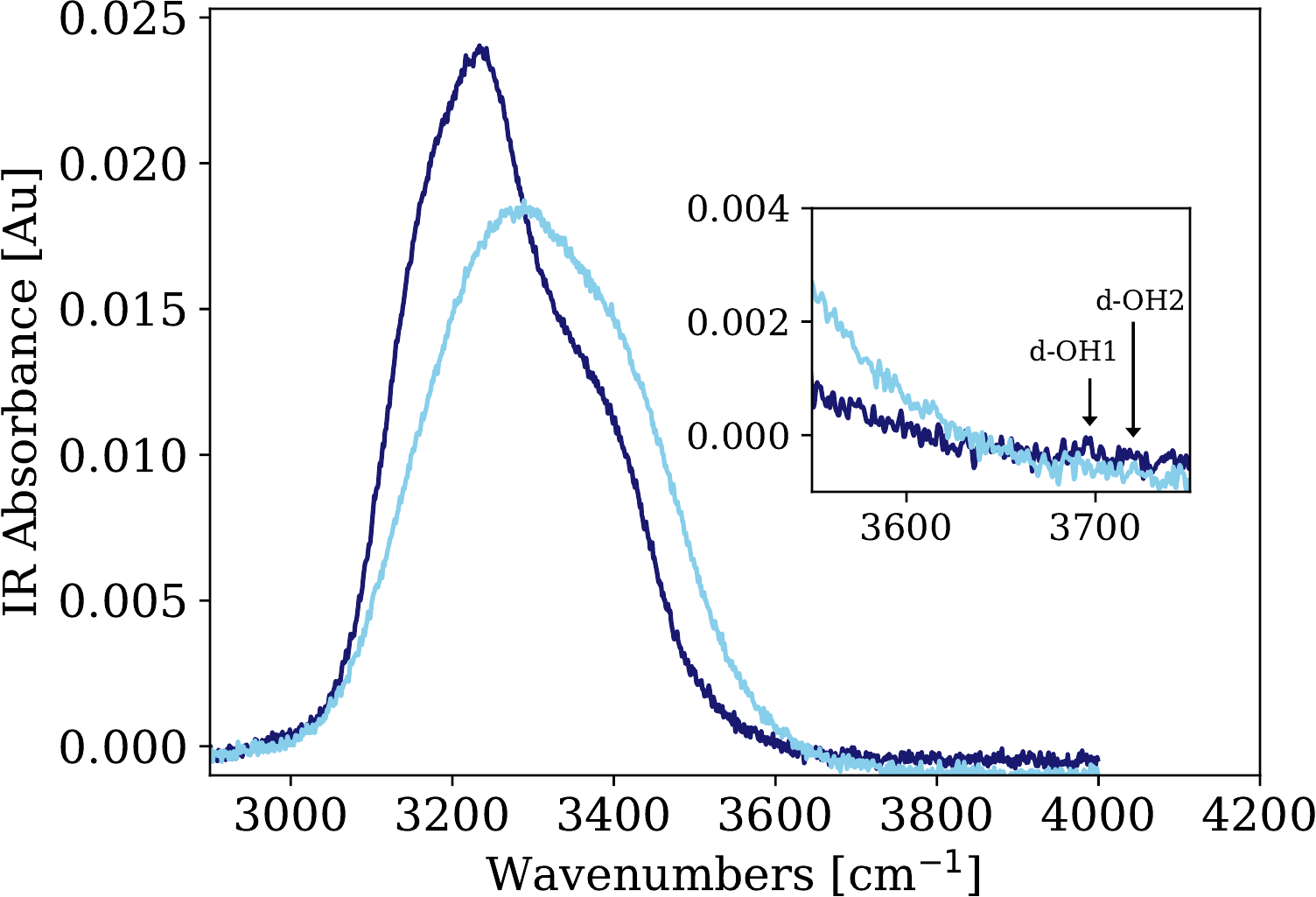} % first figure itself
    \end{minipage}\hfill
    \begin{minipage}{0.5\textwidth}
        \centering
        \includegraphics[width=0.97\textwidth]{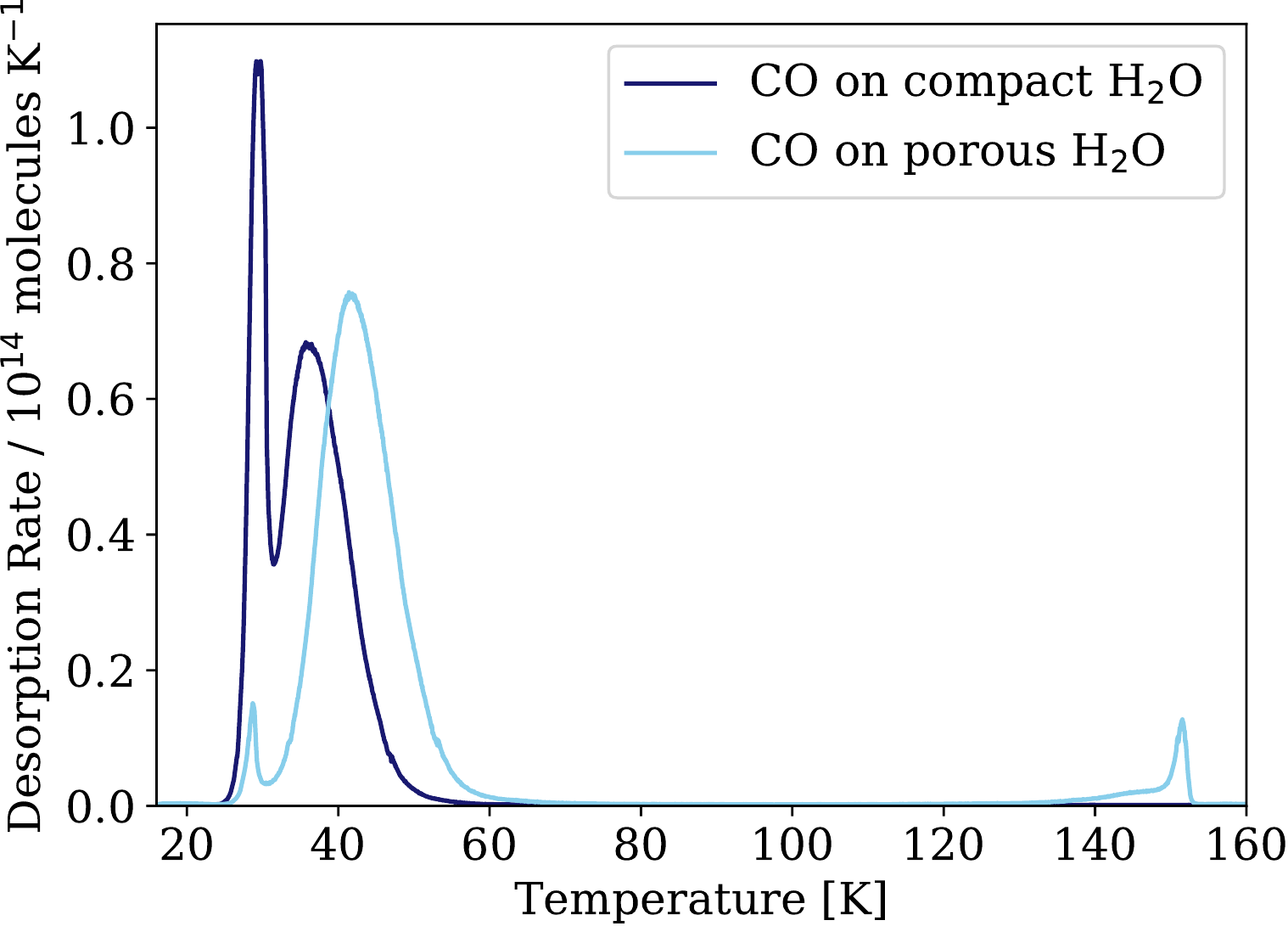} % second figure itself
    \end{minipage}
		\caption{Left: FTIR data for one representative porous and one compact H$_{2}$O ice substrate, with an inset centered on where the dangling O-H bond features would be located (d-O-H1 at $\sim$3697 cm$^{-1}$, d-O-H2 at $\sim$3720 cm$^{-1}$). The degree of porosity cannot be determined from the dangling O-H bond feature as it is not visible in the spectra. Right: TPD curves of CO on compact and porous H$_{2}$O substrates. There is a CO desorption peak around the H$_{2}$O desorption temperature for the porous substrate experiment, but not for the compact substrate experiment.}
		\label{fig:figure9}
\end{figure*}

%\section{Arrhenius Plots}
%\counterwithin{figure}{section}

\begin{figure*}[t]
	        \centering
		\includegraphics[width=0.98\textwidth]{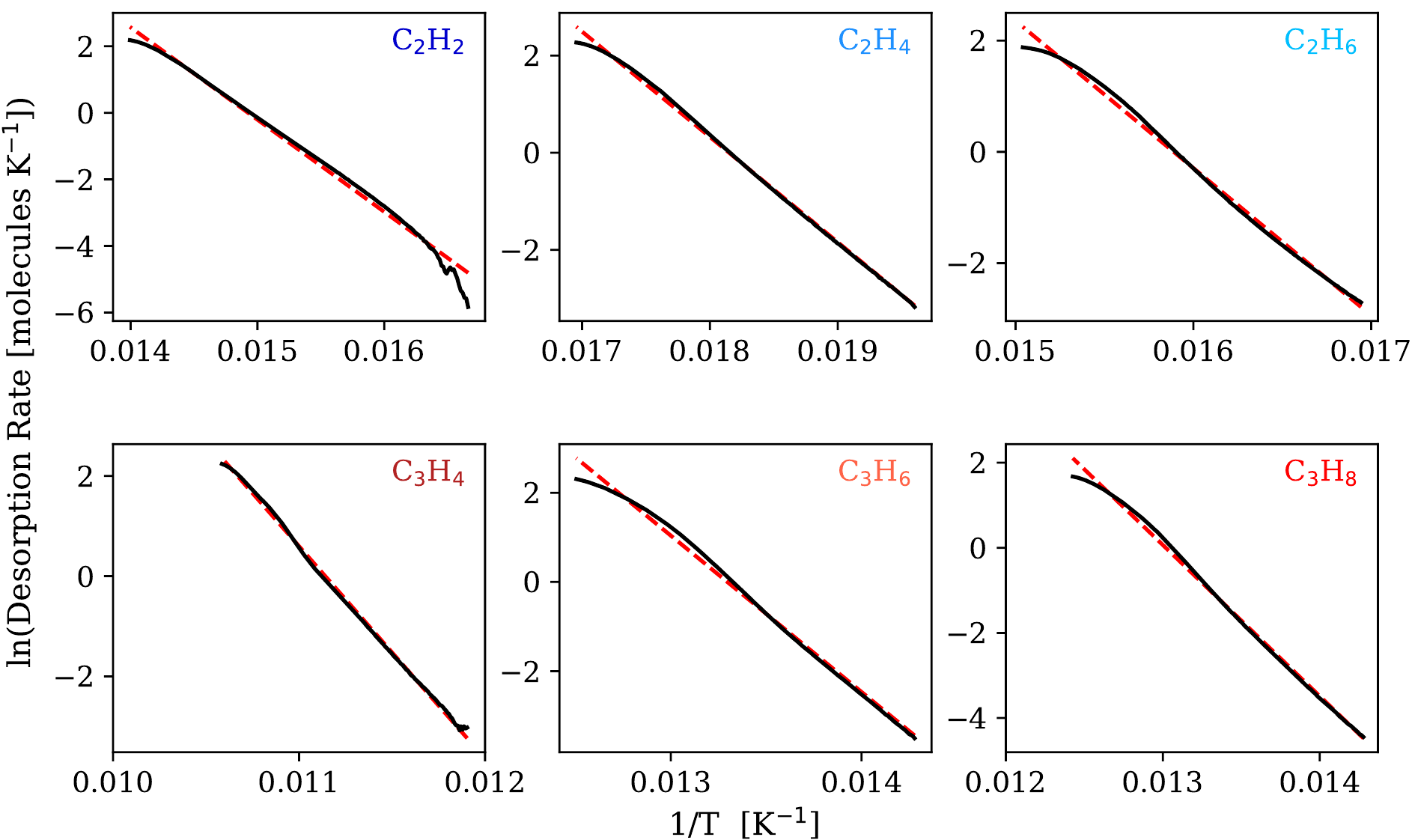}
		\vspace{3mm}
		\caption{Arrhenius plots for all pure TPDs with fits to determine the binding energies $E_{b}$ and pre-exponential factors $\nu$. The desorption rate data are represented by the black lines, while the fits are represented by the dashed red lines.}
		\label{fig:figure10}
\end{figure*}

\begin{figure*}[t]
	        \centering
    \begin{minipage}{0.48\textwidth}
        \centering
        \includegraphics[width=1.03\textwidth]{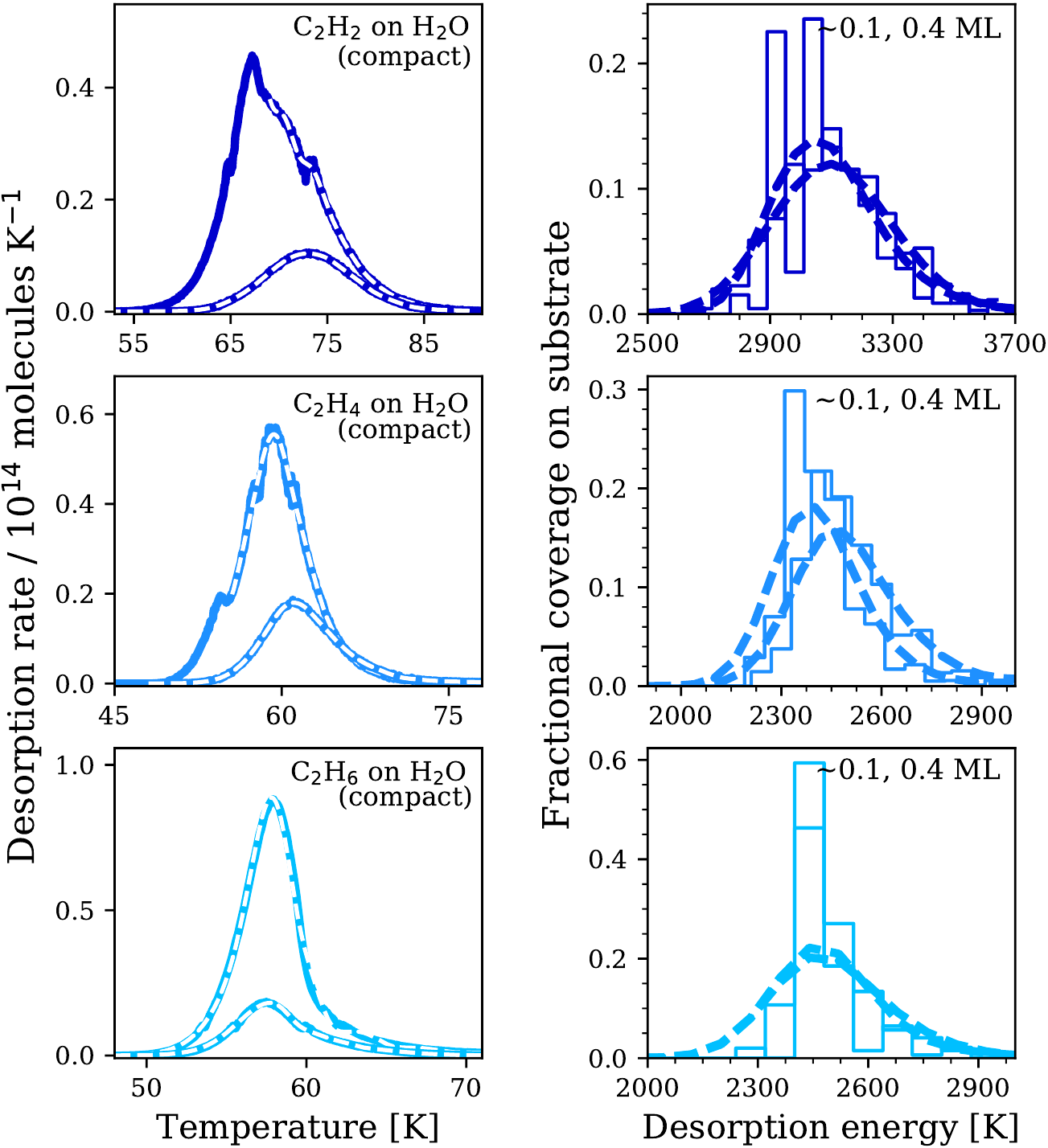} % first figure itself
    \end{minipage}\hfill
    \begin{minipage}{0.5\textwidth}
        \centering
        \includegraphics[width=0.99\textwidth]{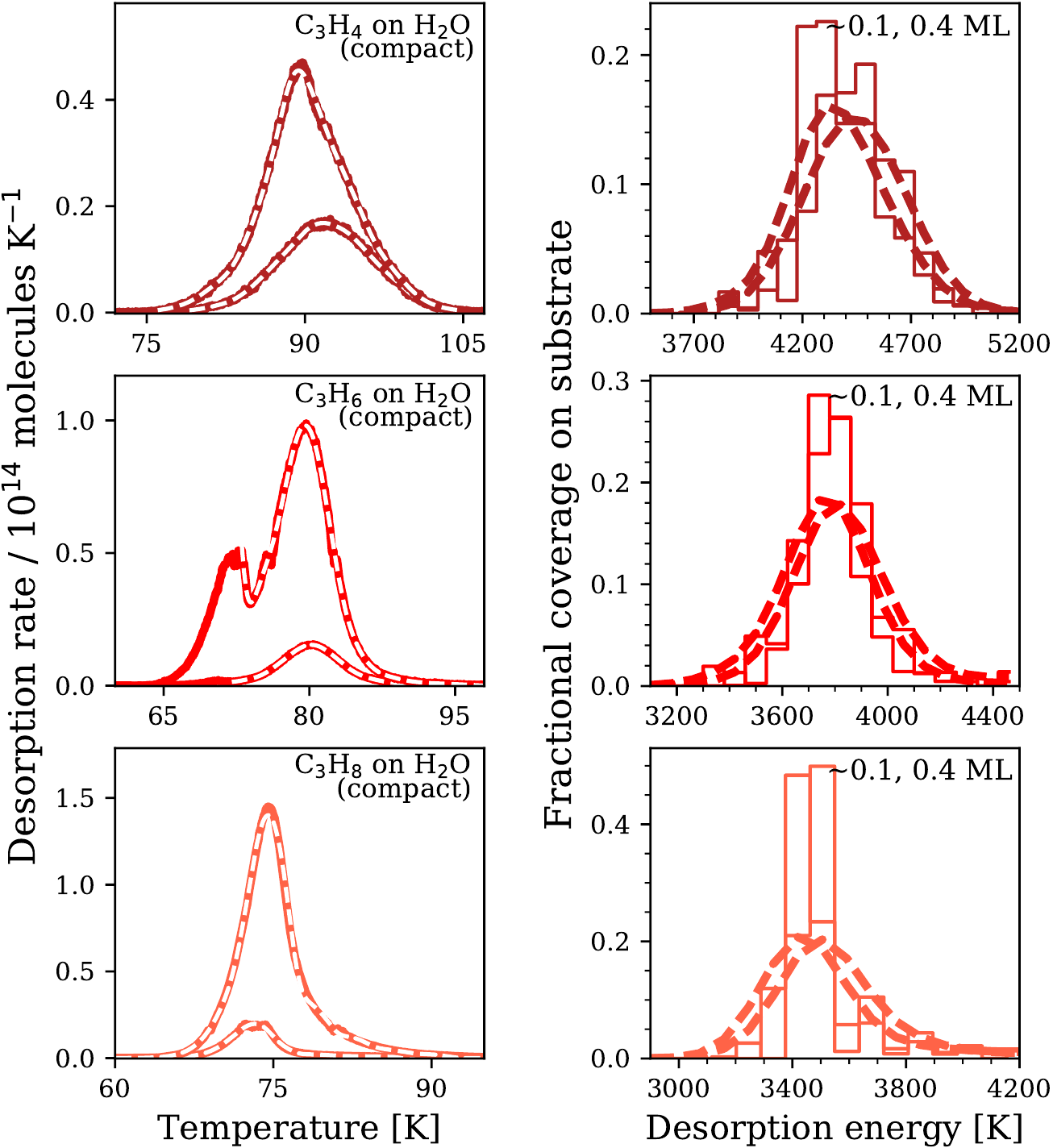} % second figure itself
    \end{minipage}
		\caption{2- and 3-carbon hydrocarbon TPD curves on compact H$_{2}$O substrates (coverages of $\sim$0.1, 0.4 ML) and corresponding binding energy distributions, similar to Fig.'s~\ref{fig:figure5} and~\ref{fig:figure6}. The variation in binding energy distribution centroids for coverages between $\sim$0.1, 0.2 (target coverage), and 0.4 ML is less than 200 K for all hydrocarbons used in this study.}
		\label{fig:figure11}
\end{figure*}




\newpage
\bibliography{mybib}
\bibliographystyle{apj}



\end{document}